\documentclass[aps,prd,twocolumn,nofootinbib,superscriptaddress]{revtex4}
\usepackage{amsmath,graphicx}
\usepackage{subfigure}

\usepackage[dvipdfm,colorlinks=true,citecolor=blue,pdfstartview=FitH]{hyperref}

\usepackage{color,framed} 

\def\bea{\begin{equation}}
\def\eea{\end{equation}}
\newcommand{\QH}{quadruply heavy}
\newcommand{\ptq}{pentaquark ${bb\bar{u}cc}$}
\newcommand{\rt}{Regge trajectory}
\newcommand{\rts}{Regge trajectories}
\newcommand{\tr}{trajectory}
\newcommand{\trs}{trajectories}
\newcommand{\bfr}{{\bf r}}
\newcommand{\bfp}{{\bf p}}
\newcommand{\bfpa}{{|\bf p|}}
\newcommand{\gev}{{\rm GeV}}

\newcommand{\cltb}{$\bar{3}_c$}
\newcommand{\cltba}{\bar{3}_c}

\begin{document}
\title{$\lambda$, $\rho$, and $\sigma$ Regge trajectories for the quadruply heavy pentaquark $bb\bar{u}cc$ in the diquark-triquark picture}
\author{Xin-Ru Liu}
\email{1170394732@qq.com}
\affiliation{School of Physics and Electronic Engineering, Shanxi Normal University, Taiyuan 030031, China}
\author{Qi Liu}
\email{18803429267@163.com}
\affiliation{School of Physics and Electronic Engineering, Shanxi Normal University, Taiyuan 030031, China}
\author{Jiao-Kai Chen}
\email{chenjk@sxnu.edu.cn, chenjkphy@outlook.com (corresponding author)}
\affiliation{School of Physics and Electronic Engineering, Shanxi Normal University, Taiyuan 030031, China}

\begin{abstract}
Systematic investigations of four series of Regge trajectories for quadruply heavy pentaquarks are still lacking. Using the diquark and triquark Regge trajectory relations, we propose the Regge trajectory relations for the quadruply heavy pentaquark ${bb\bar{u}cc}$: $M=2m_{b}+2m_{c}+m_{u}+5C/2 +\beta_{x_{\lambda}}(x_{\lambda}+c_{0x_{\lambda}})^{2/3}
+\beta_{x_{\rho_1}}(x_{\rho_1}+c_{0x_{\rho_1}})^{2/3}+\beta_{x_{\rho_2}}\sqrt{x_{\rho_2}+c_{0x_{\rho_2}}}
+\beta_{x_{\sigma}}(x_{\sigma}+c_{0x_{\sigma}})^{2/3}$. Four series of Regge trajectories, namely the $\lambda$-, $\rho_1$-, $\rho_2$-, and $\sigma$-trajectories, are investigated.
We demonstrate that accounting for the internal structure and substructure of pentaquarks is indispensable for constructing the $\rho_1$-, $\rho_2$-, and $\sigma$-trajectories; without such structural considerations, functional form and trajectory parameters can only be obtained via pure fitting against theoretical or experimental data.
We further prove that the Regge trajectories of diquark 1, triquark, and diquark 2 (embedded within the triquark) do not correspond one to one to the $\rho_1$-, $\rho_2$-, and $\sigma$-trajectories. Nevertheless, these trajectories govern the behaviors of the respective $\rho_1$-, $\rho_2$-, and $\sigma$-trajectories.
For both configurations $(bb)(\bar{u}(cc))$ and $(cc)(\bar{u}(bb))$, the $\lambda$-, $\rho_1$-, and $\sigma$-trajectories exhibit behavior of $M{\sim}x^{2/3}$ ($x=n_{r_1},n_{r_3},l_1,l_3,N_{r},L$), whereas the $\rho_2$-trajectories exhibit behavior of $M{\sim}\sqrt{x}$ ($x=n_{r_2},\,l_2$). The functional behavior of Regge trajectories for diquarks and triquark offers guidance for fitting the pentaquark Regge trajectories.
Additionally, we provide rough estimates for spin-averaged masses of the $\lambda$-, $\rho_1$-, $\rho_2$-, and $\sigma$-excited states.
\end{abstract}

\maketitle


\section{Introduction}

Tremendous progress has been achieved in both experimental and theoretical explorations of pentaquarks \cite{takahashi:pdg2026,Brambilla:2019esw,Jaffe:2004ph,Liu:2019zoy} since Gell-Mann suggested the possible existence of multiquark states in 1964 \cite{Gell-Mann:1964ewy}.
However, none of pentaquark candidates can survive scrutiny of additional experimental data until a study of $\Lambda^0_b\to J/{\psi}K^-p$ decays by LHCb \cite{LHCb:2015yax}. In 2015, LHCb Collaboration reported observations of exotic structures in the $J/{\Psi}p$ channel, which is referred to as charmonium-pentaquark states, in $\Lambda^0_b\to J/{\psi}K^-p$ decays.
In 2019 \cite{LHCb:2019kea}, the peak previously seen at 4450 {\rm MeV} was resolved into two narrowly separated resonances, $P_c(4440)$ and $P_c(4457)$. In addition, a new narrow resonance, $P_c(4312)$, was also discovered.
The LHCb Collaboration discovered $P_{cs}(4459)$ in $\Xi^-_b{\to}J/{\Psi}\Lambda{K^-}$ decays in 2020 \cite{LHCb:2020jpq}, reported $P_c(4337)$ in $B^0_s{\to}J/{\Psi}p\bar{p}$ decays in 2021 \cite{LHCb:2021chn}, and observed $P_{cs}(4338)$ in $B^-{\to}J/{\Psi}\Lambda\bar{p}$ decays in 2022 \cite{LHCb:2022ogu}. In 2025, Belle and Belle-II collaborations found evidence of the $P_{c\bar{c}s}(4459)^0$ state using the first observations of $\Upsilon(1S, 2S)$ inclusive decays to $J/\Psi\Lambda$ \cite{Belle:2025pey}.

Pentaquarks have been discussed in various pictures, such as the diquark-triquark picture \cite{Karliner:2003dt,Lebed:2015tna,Zhu:2015bba}, the soliton picture \cite{Scoccola:2015nia}, the diquark-diquark-antiquark picture \cite{Jaffe:2003sg,Maiani:2015vwa,Wang:2026thx,Anisovich:2015cia}, the meson-baryon picture \cite{Karliner:2015ina,Chen:2015loa,Ortega:2016syt,Roca:2015dva,
Burns:2015dwa,Jing:2025iqs,Song:2025yut,Mutuk:2026zxp}, the colored meson-baryon picture \cite{Mironov:2015ica,Takeuchi:2016ejt,Deng:2016rus,Santopinto:2016pkp,Wu:2017weo}, the five-quark picture \cite{An:2022fvs,An:2021vwi}, the four-quark-antiquark picture \cite{Genovese:1997tm,Stancu:2003if}, the quark-meson-diquark picture \cite{Garcilazo:2022kra}, among other things.

In the present work, we systematically investigate four series of {\rts} for the {\ptq} in the diquark-triquark picture.
The {\rt} is one of the effective approaches widely used in the study of hadron spectra
\cite{Burns:2010qq,Regge:1959mz,Chew:1962eu,Nambu:1978bd,Gross:2022hyw,Brodsky:2006uq,Nielsen:2018uyn,
brau:04bs,Brisudova:1999ut,Guo:2008he,Irving:1977ea,Collins:1971ff,
Inopin:1999nf,Afonin:2014nya,MartinContreras:2020cyg,Sergeenko:1994ck,Veseli:1996gy,
Wilczek:2004im,Selem:2006nd,Sonnenschein:2018fph,MartinContreras:2023oqs,Roper:2024ovj,Jakhad:2025zos,
Chen:2018hnx,Chen:2018bbr,Chen:2018nnr}. 
To date, only a small number of works have employed the {\rt} approach to the discussion of pentaquarks \cite{Sindhu:2023oqo,Ghosh:2017cck,Song:2025cla}. In Refs. \cite{Sindhu:2023oqo,Ghosh:2017cck}, only the $\lambda$-{\trs} were discussed. In Ref. \cite{Song:2025cla}, we discussed four series of {\rts} for the fully heavy pentaquark ${P}_{cc\overline{c}bb }$. Systematical studies on four series of {\rts} for the quadruply heavy pentaquarks remain absent. First, we propose the {\rt} relations for the {\ptq} by employing the diquark and triquark {\rts} \cite{Feng:2023txx,Liu:2026npi}. Then, four series of {\rts} are investigated by employing the newly proposed relations. Lastly, we roughly estimate the masses of the $\lambda$-, $\rho_1$-, $\rho_2$-, and $\sigma$-excited states.

The paper is organized as follows. In Sec. \ref{sec:rgr}, the {\rt} relations for the quadruply heavy pentaquarks are proposed. In Sec. \ref{sec:rts}, four series of masses and four series of {\rts} are presented. The conclusions are summarized in Sec. \ref{sec:conc}.

\section{{\rt} relations for the {\QH} pentaquark ${bb\bar{u}cc}$}\label{sec:rgr}
In this section, by utilizing the diquark {\rts} \cite{Feng:2023txx} and the triquark {\rts} \cite{Liu:2026npi}, we propose the {\rt} relations for the {\QH} {\ptq}, which can be employed to discuss the $\lambda$-, $\rho$-, and $\sigma$-{\trs}.

\subsection{Preliminary}\label{subsec:prelim}

\begin{figure}[!phtb]
\centering
\includegraphics[width=0.26\textheight]{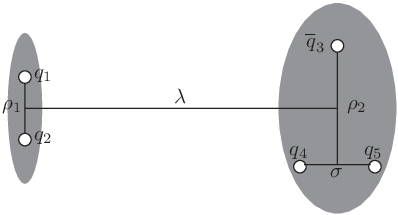}
\caption{Schematic diagram of a pentaquark in the diquark-triquark picture. The left gray part represents the diquark $1$, which is composed of quarks $q_1$ and $q_2$. The right gray part denotes a triquark, composed of one antiquark $\bar{q}_3$ and diquark $2$. Diquark $2$ contains quarks $q_4$ and $q_5$. The circles stand for quarks and antiquarks.}\label{fig:pr}
\end{figure}

In the diquark-triquark picture, a pentaquark consists of one diquark and one triquark (see Fig. \ref{fig:pr}). $\rho_1$ and $\sigma$ separate the quarks within diquark $1$ and diquark $2$, respectively. $\lambda$ corresponds to the separation between diquark $1$ and the triquark, while $\rho_2$ corresponds to the separation between the antiquark $\bar{q}_3$ and diquark $2$. Four excited modes exist: the $\rho_1$-mode describes radial and orbital excitation inside diquark $1$; the $\sigma$-mode involves excitation within diquark $2$; the $\lambda$-mode involves radial or orbital excitation between diquark $1$ and the triquark; and the $\rho_2$-mode involves excitation between the antiquark $\bar{q}_3$ and diquark $2$. Accordingly, four series of {\rts} emerge: one series of $\lambda$-{\trs}, two series of $\rho$-{\trs}, and one series of $\sigma$-{\trs}.

A diquark $(q_1q_2)$ can couple to two irreducible color representations: $3_c\otimes3_c=\cltba\oplus{6}_c$. The $\bar{3}_c$ is the attractive channel, while in the $6_c$ representation, the internal interaction between the $q_1q_2$ pair is repulsive.
Some studies take both $\cltba$ and ${6}_c$ into account (see, e.g., Refs. \cite{Maiani:2019lpu,Berwein:2024ztx}), whereas other works retain only the $\cltba$ channel (see, .e.g., Refs. \cite{Brodsky:2014xia,Galkin:2023wox}).
Following Refs. \cite{Brodsky:2014xia,Galkin:2023wox}, only the $\bar{3}_c$ diquark is considered in the present work. Likewise, the triquark $(\bar{q}_3(q_4q_5))$ considered herein is a ${3}_c$ bound state, composed of a diquark $(q_4q_5)$ in the $\cltba$ representation and an antiquark $\bar{q}_3$ that also is in the $\cltba$ representation. In the triquark-diquark model, the color-singlet pentaquarks under consideration are composed of a $\bar{3}_c$ diquark and a $3_c$ triquark \cite{Lebed:2015tna}.

In the diquark-triquark picture, the pentaquark state is denoted as
\bea\label{tetnot}
\left((q_1q_2)^{{\bar{3}_c}}_{n_1^{2s_1+1}l_{1j_1}}
\left(\bar{q}_3(q_4q_5)^{{\bar{3}_c}}_{n_3^{2s_3+1}l_{3j_3}}\right)
^{{{3}_c}}_{n_2^{2s_{2}+1}l_{2j_2}}\right)^{1_c}_{N^{2S+1}L_{J}},
\eea
where {\cltb} denotes the color antitriplet state of the diquark, and $1_c$ represents the color singlet state of the pentaquark. [The superscript $1_c$ is often omitted, as the observed pentaquarks are colorless.]
The notation in Eq. (\ref{tetnot}) can also be written as $|n_1^{2s_1+1}l_{1j_1},n_2^{2s_2+1}l_{2j_2},n_3^{2s_{3}+1}l_{3j_3},N^{2S+1}L_{J}\rangle$.
The diquark $(q_1q_2)$ is either $\{q_1q_2\}$ or $[q_1q_2]$, where $\{q_1q_2\}$ and $[q_1q_2]$ represent the permutation symmetric and antisymmetric flavor wave functions, respectively. $N=N_{r}+1$, where $N_{r}=0,\,1,\,\cdots$. $n_{1,2,3}=n_{r_{1,2,3}}+1$, where $n_{r_{1,2,3}}=0,\,1,\,\cdots$. $N_{r}$, $n_{r_1}$, $n_{r_2}$ and $n_{r_3}$ are the radial quantum numbers of the pentaquark, diquark $1$, triquark, and diquark 2, respectively.
\begin{align}
\vec{J}=&\vec{L}+\vec{S},\; \vec{S}=\vec{j}_1+\vec{j}_2,\nonumber\\
\vec{j}_1=&\vec{s}_1+\vec{l}_1,\; \vec{s}_1=\vec{s}_{q_1}+\vec{s}_{q_2},\nonumber\\
\vec{j}_2=&\vec{s}_{2}+\vec{l}_2,\;\vec{s}_{2}=\vec{j}_3+\vec{s}_{\bar{q}_3},\nonumber\\
\vec{j}_3=&\vec{s}_3+\vec{l}_3,\; \vec{s}_3=\vec{s}_{q_4}+\vec{s}_{q_5}.
\end{align}
$\vec{J}$, $\vec{j}_1$, $\vec{j}_2$ and $\vec{j}_3$ are the total angular momentums of pentaquark, diquark $1$, triquark, and diquark $2$, respectively. $L$, $l_1$, $l_{2}$ and $l_3$ are the orbital quantum numbers of pentaquark, diquark $1$, triquark, and diquark $2$, respectively.
$\vec{S}$ is the total spin of diquark $1$ and triquark; $\vec{s}_{1}$ is the total spin of quarks in diquark $1$; $\vec{s}_{2}$ is the total spin of antiquark $3$ and diquark $2$; $\vec{s}_{3}$ is the total spin of quarks in diquark $2$.
$\vec{s}_{q_1}$, $\vec{s}_{q_2}$, $\vec{s}_{\bar{q}_3}$, $\vec{s}_{q_4}$, and $\vec{s}_{q_5}$ are spins of quark 1, quark 2, antiquark 3, quark 4, and quark 5, respectively.

In the diquark-triquark picture, the quadruply heavy {\ptq} has three configurations: $(bb)(\bar{u}(cc))$, $(cc)(\bar{u}(bb))$, and $(bc)(\bar{u}(bc))$. Due to mode mixing in the $(bc)(\bar{u}(bc))$ configuration, the corresponding {\rts} are complicated. This configuration is therefore neglected in the present study.

\subsection{Spinless Salpeter equation}
The spinless Salpeter equation \cite{Godfrey:1985xj,Capstick:1986ter,Ferretti:2019zyh,Bedolla:2019zwg,Durand:1981my,Durand:1983bg,Lichtenberg:1982jp,Jacobs:1986gv} takes the form
\begin{eqnarray}\label{qsse}
M\Psi_{d,t,p}({\bfr})=\left(\omega_1+\omega_2\right)\Psi_{d,t,p}({\bfr})+V_{d,t,p}\Psi_{d,t,p}({\bfr}),
\end{eqnarray}
where $M$ is the bound state mass (diquark, triquark, or pentaquark). $\Psi_{d,t,p}({\bfr})$ are the diquark wave function, triquark wave function, and pentaquark wave function, respectively. $\omega_1$ is the relativistic energy of constituent $1$ (quark, diquark, or antiquark), while $\omega_2$ is that of constituent $2$ (quark, diquark, or triquark),
\bea\label{omega}
\omega_i=\sqrt{m_i^2+{\bf p}^2}=\sqrt{m_i^2-\Delta}\;\; (i=1,2).
\eea
$m_1$ and $m_2$ are the effective masses of constituent $1$ and $2$, respectively.

Following Refs. \cite{Ferretti:2019zyh,Bedolla:2019zwg,Ferretti:2011zz,Eichten:1974af}, we adopt the potential
\begin{align}\label{potv}
V_{d,t,p}&=-\frac{3}{4}\left[V_c+{\sigma}r+C\right]
\left({\bf{F}_i}\cdot{\bf{F}_j}\right)_{d,t,p},
\end{align}
where $V_c\propto{1/r}$ is a color Coulomb potential or a Coulomb-like potential due to one-gluon exchange. $\sigma$ is the string tension. $C$ is a fundamental parameter \cite{Gromes:1981cb,Lucha:1991vn}. The part in the brackets is the Cornell potential \cite{Eichten:1974af}. ${\bf{F}_i}\cdot{\bf{F}_j}$ is the color-Casimir,
\bea\label{mrcc}
\langle{(\bf{F}_i}\cdot{\bf{F}_j})_{d,t}\rangle=-\frac{2}{3},\quad
\langle{(\bf{F}_i}\cdot{\bf{F}_j})_{p}\rangle=-\frac{4}{3}.
\eea

The Regge-trajectory approach has two ways to discuss spin-dependent masses. One is to calculate the spin-averaged masses using the {\rt} formula at first, and then obtain the spin-dependent corrections by calculating the spin-dependent terms \cite{Jakhad:2025zos}; the other is to directly calculate the spin-dependent masses using the {\rt} formulas \cite{Guo:2008he,Sonnenschein:2018fph,MartinContreras:2020cyg,Roper:2024ovj,Burns:2010qq}. Both ways can yield reasonable results. When calculating the spin-averaged masses, no extra errors are introduced within the former way although neglecting the spin-dependent terms is a drastic simplification.  
In the present work, the interaction potential considered includes only the central terms [Eq. (\ref{potv})], and the pentaquark masses calculated are spin averaged. The contact term, the spin-orbit term, and the Thomas precession term are relativistic corrections and are small compared with the Cornell potential. These contributions do not alter the {\rt} behavior. We concentrate on the {\rt} behavior for pentaquarks and therefore neglect these spin-dependent effects.

\subsection{{\rt} relations for heavy-heavy and heavy-light systems}
For heavy-heavy systems, since $m_{1},m_2{\gg}{\bfpa}$, Eq. (\ref{qsse}) reduces to
\begin{eqnarray}\label{qssenrr}
M\Psi_{d,t,p}({\bfr})=\left[(m_1+m_2)+\frac{{\bfp}^2}{2\mu}+V_{d,t,p}\right]
\Psi_{d,t,p}({\bfr}),
\end{eqnarray}
where
\bea\label{rdmu}
\mu=\frac{m_1m_2}{m_1+m_2}.
\eea
By employing the Bohr-Sommerfeld quantization approach \cite{brau:04bs} and using Eqs. (\ref{potv}) and (\ref{qssenrr}), we obtain the parametrized relation  \cite{Chen:2022flh,Chen:2021kfw}
\begin{align}\label{massform}
M=&m_R+\beta_x(x+c_{0x})^{2/3},\nonumber\\
x=&l_1,\,l_3,\,n_{r_1},\,n_{r_3},\,L,\,N_r.
\end{align}
The coefficients $\beta_x$ and $m_R$ are
\bea\label{parabm}
\beta_x=c_{fx}c_xc_c,\quad m_R=m_1+m_2+C',
\eea
where $c_{x}$ and $c_c$ are
\begin{align}\label{cxcons}
&c_c=\left(\frac{\sigma'^2}{\mu}\right)^{1/3},\quad
c_{l_1,l_2,l_3,L}=\frac{3}{2},\nonumber\\ &c_{n_{r_1},n_{r_2},n_{r_3},N_r}=\frac{\left(3\pi\right)^{2/3}}{2},
\end{align}
\begin{align}\label{cprime}
C'&=\left\{\begin{array}{cc}
C/2, & \text{diquarks,\,triquarks}, \\
C, & \text{pentaquarks}.
\end{array}\right. \nonumber\\
\sigma'&=\left\{\begin{array}{cc}
\sigma/2, & \text{diquarks,\,triquarks}, \\
\sigma, & \text{pentaquarks}.
\end{array}\right.
\end{align}
$c_{fx}$ are theoretically unity but treated as free fitting parameters in practical calculations.
In Eq. (\ref{massform}), $m_1$, $m_2$, $c_x$ and $\sigma$ are universal for heavy-heavy systems. $c_{0x}$ vary with different {\rts}.

For heavy-light systems with $m_1\to\infty$ and $m_2\to0$, Eq. (\ref{qsse}) simplifies to
\begin{eqnarray}\label{qssenr}
M\Psi_{d}({\bfr})=\left[m_1+{\bfpa}+V_{d}\right]\Psi_{d}({\bfr}).
\end{eqnarray}
By employing the Bohr-Sommerfeld quantization approach \cite{brau:04bs} and using Eq. (\ref{qssenr}), the parametrized formula can be obtained \cite{Chen:2022flh,Chen:2021kfw}
\bea\label{rtmeson}
M=m_R+\beta_x\sqrt{x+c_{0x}}\;(x=l_2,\,n_{r_2}).
\eea
$\beta_x$ is in Eq. (\ref{parabm}), while the corresponding coefficients read
\bea\label{cxcons}
c_{c}=\sqrt{\sigma'},\quad c_{l_2}=2,\quad c_{n_{r_2}}=\sqrt{2\pi}.
\eea
For heavy-light systems, the common choice of $m_R$ is \cite{Selem:2006nd,Chen:2021kfw,Veseli:1996gy}
\bea\label{mrm1}
m_R=m_1.
\eea

The conventional {\rt}, given by Eqs. (\ref{rtmeson}) and (\ref{mrm1}), is obtained in the limit $m_1\to\infty$ and $m_2\to0$.
In Ref. \cite{Chen:2023cws}, we proposed two modified formulas that incorporate the light constituent's mass, enabling an unified description of both heavy-light mesons and heavy-light diquarks.
The first formula is Eq. (\ref{rtmeson}) with $m_R$ in (\ref{parabm}), where $m_2$ denotes the mass of the light constituent. The second reads
\bea\label{mrtf}
M=m_R+\sqrt{\beta_x^2(x+c_{0x})+\kappa_{x}m^{3/2}_2(x+c_{0x})^{1/4}},
\eea
valid under the condition $m_2{\ll}M$, with
\bea\label{mrfp}
m_R=m_1+C',\quad \kappa_x=\frac{4}{3}\sqrt{{\pi}\beta_x}\,,
\eea
where $\beta_x$ is in (\ref{parabm}).
Equation (\ref{rtmeson}) with (\ref{parabm}) generalizes the relation \cite{Afonin:2014nya}
\bea\label{afoninequ}
M=m_1+m_2+\sqrt{a(n_r+{\alpha}l+b)}
\eea
and the formula \cite{Chen:2022flh}
\bea\label{rmfnpb}
(M-m_1-m_2-C)^2=\alpha_x(x+c_0)^{\gamma}
\eea
whereas (\ref{mrtf}) with (\ref{mrfp}) is based on the results in Refs. \cite{Selem:2006nd,Sonnenschein:2018fph}.
As $m_2=0$, these two modified formulas, Eq. (\ref{rtmeson}) with (\ref{parabm}) and Eq. (\ref{mrtf}) with (\ref{mrfp}), become identical. As $m_2=0$ and $C$ is neglected, these two modified formulas reduce to the usual {\rt} formula for heavy-light systems, namely Eqs. (\ref{rtmeson}) and (\ref{mrm1}).
Although they give different behavior of $m_2$, Eq. (\ref{rtmeson}) with (\ref{parabm}) and Eq. (\ref{mrtf}) with (\ref{mrfp}) produce consistent results for $l,\,n_r<10$ and have the same behavior $M{\sim}x^{1/2}$ \cite{Chen:2023cws}.

In the loaded flux tube model, the {\rt} formula for heavy-light systems reads $M=m_1+\sqrt{{\sigma}L/2}$ in the limit $m_1\to\infty$ and $m_2\to0$. For finite $m_2\ne0$, $M=m_1+\sqrt{{\sigma}L/2}+2^{1/4}{\kappa}m_2^{3/2}L^{-1/4}$ is given in Ref. \cite{Selem:2006nd}, where the correction term is determined via computer simulations. In the relativistic flux tube model \cite{Jakhad:2025zos}, the small mass correction term is derived under condition $m_1{\gg}m_2$, with $m_c=1.37568$ {\gev} and $m_u=0.39995$ {\gev}. To the best of our knowledge, in the potential model, no {\rt} formula can be analytically derived from Eq. (\ref{qsse}) for large $m_1$ and small $m_2$ without approximation, owing to its complexity. Therefore, we first derive a simple formula by taking the limit $m_1\to\infty$ and $m_2\to0$, and then modify it by incorporating the mass of light constituent. Equation (\ref{rtmeson}) with (\ref{parabm}) and Eq. (\ref{mrtf}) with (\ref{mrfp}) are obtained under the condition $M{\gg}m_2$, namely $m_1{\gg}m_2$. However, the given parameters $m_c=1.55$ {\gev} and $m_u=0.33$ {\gev} hardly seem to satisfy this "much larger than" criterion. In practice, the fitted results from the modified formula are superior to those from the unmodified one. The modified relation yields results that are consistent with other theoretical predictions for heavy mesons and heavy diquarks; see Ref. \cite{Chen:2023cws} for further details. It is also expected to be applicable to doubly heavy triqurks $(\bar{u}(QQ))$.

\begin{table}[!phtb]
\caption{The coefficients for heavy-heavy systems (HHS) and heavy-light systems (HLS).}  \label{tab:eparam}
\centering
\begin{tabular*}{0.47\textwidth}{@{\extracolsep{\fill}}ccc@{}}
\hline\hline
                   & HHS &  HLS   \\
\hline
$\nu$    & $2/3$ & $1/2$    \\
$c_c$    & $\left({\sigma'^2}/{\mu}\right)^{1/3}$    & $\sqrt{\sigma'}$  \\
$c_{l_1,l_2,l_3,L}$    & $3/2$ & $2$   \\
$c_{n_{r_1},n_{r_2},n_{r_3},N_r}$ & ${\left(3\pi\right)^{2/3}}/{2}$      & $\sqrt{2\pi}$   \\
\hline
\hline
\end{tabular*}
\end{table}

When adopting Eq. (\ref{rtmeson}) with (\ref{parabm}) to discuss heavy-light systems, we unify Eqs. (\ref{massform}), (\ref{rtmeson}), and (\ref{parabm}) into a general {\rt} ansatz \cite{Chen:2022flh,Xie:2024lfo}
\begin{align}\label{massfinal}
M=&m_R+\beta_x(x+c_{0x})^{\nu},\nonumber\\
m_R=&m_1+m_2+C',\quad \beta_x=c_{fx}c_xc_{c},
\end{align}
where $x=l_1,\,l_2,\,l_3,\,n_{r_1},\,n_{r_2},\,n_{r_3},\,L,\,N_r$. The exponents ${\nu}$ and coefficients $c_x$ and $c_{c}$ are tabulated in Table \ref{tab:eparam}. $c_{fx}$ are theoretically equal to unity and treated as free fitting parameters in practical computations. $c_{0x}$ varies with different {\rts}. Equation (\ref{massfinal}) can be employed to discuss various systems covering both heavy-heavy systems and heavy-light systems, including diquarks, mesons, baryons, triquarks, tetraquarks, pentaquarks, and hexaquarks \cite{Chen:2023djq,Chen:2023ngj,Chen:2023web,
Chen:2025fyh,Song:2025cla,Song:2024bkj,Liu:2026npi}.

\subsection{{\rt} relations for the {\QH} {\ptq}}

For configurations $(bb)(\bar{u}(cc))$ and $(cc)(\bar{u}(bb))$, the pentaquark as well as diquarks $(bb)$ and $(cc)$ are heavy-heavy systems, while triquark $(\bar{u}(QQ))$ (with $QQ=cc,\,bb$) is a heavy-light system. Using Eq. (\ref{massfinal}) and Table \ref{tab:eparam}, we obtain the {\rt} relation for the {\QH} {\ptq}:
\begin{align}\label{ppt2q}
M&=m_{R_{\lambda}}+\beta_{x_{\lambda}}(x_{\lambda}+c_{0x_{\lambda}})^{2/3}\;
(x_{\lambda}=L,\,N_{r}),\nonumber\\
m_{R_{\lambda}}&=M_{d_1}+M_{t}+C,\nonumber\\
M_{d_1}&=m_{R_{\rho_1}}+\beta_{x_{\rho_1}}(x_{\rho_1}+c_{0x_{\rho_1}})^{2/3}\;
(x_{\rho_1}=l_1,\,n_{r_1}),\nonumber\\
m_{R_{\rho_1}}&=m_{q_1}+m_{q_2}+C/2,\nonumber\\
M_{t}&=m_{R_{\rho_2}}+\beta_{x_{\rho_2}}\sqrt{x_{\rho_2}+c_{0x_{\rho_2}}}\;
(x_{\rho_2}=l_2,\,n_{r_2}),\nonumber\\
m_{R_{\rho_2}}&=M_{d_2}+m_{u}+C/2,\nonumber\\
M_{d_2}&=m_{R_{\sigma}}+\beta_{x_{\sigma}}(x_{\sigma}+c_{0x_{\sigma}})^{2/3}\;
(x_{\sigma}=l_3,\,n_{r_3}),\nonumber\\
m_{R_{\sigma}}&=m_{q_4}+m_{q_5}+C/2,
\end{align}
where
\begin{align}\label{pppa2qQ}
\mu_{\lambda}&=\frac{M_{d_1}M_{t}}{M_{d_1}+M_{t}},\;
\mu_{\rho_1}=\frac{m_{q_1}m_{q_2}}{m_{q_1}+m_{q_2}},\nonumber\\
\beta_{L}&=\frac{3}{2}\left(\frac{\sigma^2}{\mu_{\lambda}}\right)^{1/3}c_{fL},\; \beta_{N_{r}}=\frac{(3\pi)^{2/3}}{2}\left(\frac{\sigma^2}{\mu_{\lambda}}\right)^{1/3}c_{fN_{r}},\nonumber\\
\beta_{l_1}&=\frac{3}{2}\left(\frac{\sigma^2}{4\mu_{\rho_1}}\right)^{1/3}c_{fl_1},\; \beta_{n_{r_1}}=\frac{(3\pi)^{2/3}}{2}\left(\frac{\sigma^2}{4\mu_{\rho_1}}\right)^{1/3}c_{fn_{r_1}},\nonumber\\
\beta_{l_2}&=\sqrt{2\sigma}c_{fl_2},\; \beta_{n_{r_2}}=\sqrt{\pi\sigma}c_{fn_{r_2}},\;
\mu_{\sigma}=\frac{m_{q_4}m_{q_5}}{m_{q_4}+m_{q_5}},\nonumber\\
\beta_{l_3}&=\frac{3}{2}\left(\frac{\sigma^2}{4\mu_{\sigma}}\right)^{1/3}c_{fl_3},\; \beta_{n_{r_3}}=\frac{(3\pi)^{2/3}}{2}\left(\frac{\sigma^2}{4\mu_{\sigma}}\right)^{1/3}c_{fn_{r_3}}.
\end{align}
Here, $M$, $M_t$, $M_{d_1}$, $M_{d_2}$, $m_{q_1}$, $m_{q_2}$, $m_{u}$, $m_{q_4}$, and $m_{q_5}$ are the masses of the pentaquark, triquark, diquark $1$, diquark $2$, quark $q_1$, quark $q_2$, quark $u$, quark $4$, and quark $5$, respectively.

According to Eqs. (\ref{ppt2q}) and (\ref{pppa2qQ}), we have
\bea\label{rgffl}
M=M_{d_1}+M_{t}+C+\beta_{x_{\lambda}}(x_{\lambda}+c_{0x_{\lambda}})^{2/3}\;(x_{\lambda}=L,\,N_{r})
\eea
when diquark $1$ and the triquark are regarded as structureless constituents. The binding energies of the pentaquark is $\epsilon=C+\beta_{x_{\lambda}}(x_{\lambda}+c_{0x_{\lambda}})^{2/3}$. Even though Eq. (\ref{rgffl}) shares an identical functional form for configurations $(bb)(\bar{u}(cc))$ and $(cc)(\bar{u}(bb))$, the values of $M_{d_1}$, $M_{t}$, $\beta_{x_{\lambda}}$, and $c_{0x_{\lambda}}$ differ between these two configurations.
When the triquark is taken as a bound state composed of a diquark and an antiquark, and diquarks $1$ and $2$ are each regarded as bound states of two heavy quarks, we have
\begin{align}\label{rtpf}
M=&2m_{b}+2m_{c}+m_{u}+\frac{5}{2}C \nonumber\\
&+\beta_{x_{\lambda}}(x_{\lambda}+c_{0x_{\lambda}})^{2/3}
+\beta_{x_{\rho_1}}(x_{\rho_1}+c_{0x_{\rho_1}})^{2/3}\nonumber\\
&+\beta_{x_{\rho_2}}\sqrt{x_{\rho_2}+c_{0x_{\rho_2}}}
+\beta_{x_{\sigma}}(x_{\sigma}+c_{0x_{\sigma}})^{2/3}.
\end{align}
For both configurations $(bb)(\bar{u}(cc))$ and $(cc)(\bar{u}(bb))$, the {\rt} relations have the same functional form shown in Eq. (\ref{rtpf}), whereas the parameters $\beta_{x_{\lambda}}$, $c_{0x_{\lambda}}$, $\beta_{x_{\rho_1}}$, $c_{0x_{\rho_1}}$, $\beta_{x_{\rho_2}}$, $c_{0x_{\rho_2}}$, $\beta_{x_{\sigma}}$, and $c_{0x_{\sigma}}$ take distinct values.

From the pentaquark {\rt} relations [Eqs. (\ref{ppt2q}) and (\ref{pppa2qQ}), or Eqs. (\ref{pppa2qQ}) and (\ref{rtpf})], it can be seen that there are four series of masses and, correspondingly, four series of {\rts} for the {\QH} {\ptq}: the $\lambda$-trajectories (with $x_{\rho_1}$, $x_{\rho_2}$, and $x_{\sigma}$ fixed); the $\rho_1$-trajectories (with $x_{\lambda}$, $x_{\rho_2}$, and $x_{\sigma}$ fixed); the $\rho_2$-trajectories (with $x_{\lambda}$, $x_{\rho_1}$, and $x_{\sigma}$ fixed); and the $\sigma$-trajectories (with $x_{\lambda}$, $x_{\rho_1}$, and $x_{\rho_2}$ fixed).

We refer to the {\rts} calculated from Eqs. (\ref{ppt2q}) and (\ref{pppa2qQ}), or from Eqs. (\ref{pppa2qQ}) and (\ref{rtpf}), as the complete forms of the {\rts}. The obtained constant and the mode under consideration are referred to the main part of the {\rts}. For example, when considering the $\rho_1$-trajectories for the configuration $(bb)(\bar{u}(cc))$, $\beta_{x_{\rho_2}}\sqrt{x_{\rho_2}+c_{0x_{\rho_2}}}$ and
$\beta_{x_{\sigma}}(x_{\sigma}+c_{0x_{\sigma}})^{2/3}$ are constants, while $\beta_{x_{\lambda}}(x_{\lambda}+c_{0x_{\lambda}})^{2/3}$ becomes a function of $x_{\rho_1}$ (through the dependence in $\beta_{x_{\lambda}}$ and $c_{0x_{\lambda}}$). Therefore, the main part of the $\rho_1$-trajectories is $\widetilde{m}_R+\beta_{x_{\rho_1}}(x_{\rho_1}+c_{0x_{\rho_1}})^{2/3}$, where $\widetilde{m}_R=2m_{b}+2m_{c}+m_{u}+{5C}/{2} +\beta_{x_{\rho_2}}\sqrt{x_{\rho_2}+c_{0x_{\rho_2}}}
+\beta_{x_{\sigma}}(x_{\sigma}+c_{0x_{\sigma}})^{2/3}$. The difference between the complete form of the $\rho_1$-{\tr} and its main part is $\beta_{x_{\lambda}}(x_{\lambda}+c_{0x_{\lambda}})^{2/3}$.

\section{{\rts} for the pentaquark ${bb\bar{u}cc}$}\label{sec:rts}

\subsection{Parameters}

The parameter values are listed in Table \ref{tab:parmv}. The values of $m_u$, $m_b$, $m_c$, $\sigma$, and $C$ are directly taken from Ref. \cite{Faustov:2021hjs}. $c_{fx}$ and $c_{0x}$ for the $\sigma$-modes are from Ref. \cite{Feng:2023txx}.
The parameters corresponding to the $\lambda$-mode are calculated via the following relations \cite{Xie:2024dfe} \footnote{In our previous work \cite{Xie:2024dfe,Song:2024bkj,Song:2025cla,Chen:2025fyh,Liu:2026vpi}, $m_0$ [$m_0=1$ {\gev}] was not included in the relations $c_{fL}=1.116 + 0.013 \mu_{\lambda}$, $c_{0L}=0.540- 0.141\mu_{\lambda}$, $c_{fN_{r}}=1.008 + 0.008\mu_{\lambda}$, and $c_{0N_{r}}=0.334 - 0.087\mu_{\lambda}$. In Ref. \cite{Liu:2026npi}, we correct this and introduce $m_0$ into Eq. (\ref{fitcfxnr}).}
\begin{align}
c_{fL}=&1.116 + 0.013\frac{\mu_{\lambda}}{m_0},\nonumber\\
 c_{0L}=&0.540- 0.141\frac{\mu_{\lambda}}{m_0}, \nonumber\\
c_{fN_{r}}=&1.008 + 0.008\frac{\mu_{\lambda}}{m_0},\nonumber\\
c_{0N_{r}}=&0.334 - 0.087\frac{\mu_{\lambda}}{m_0},\label{fitcfxnr}
\end{align}
where $m_0=1$ {\gev}. $\mu_{\lambda}$ is the reduced mass defined in Eq. (\ref{pppa2qQ}). The parameters for the $\rho_2$-mode are calculated by the following relations \cite{Liu:2026npi}:
\begin{align}\label{apprelA1}
{c}_{fn_{r_2}}=&\left(\frac{m_B}{m_c+m_u}\right)^{0.3}\left(1.4115- 0.2835 \frac{m_B}{m_0} \right. \nonumber\\
&   \left. + 0.03927 \left(\frac{m_B}{m_0}\right)^2 - 0.00192 \left(\frac{m_B}{m_0}\right)^3\right),\nonumber\\
{c}_{0n_{r_2}}=&\frac{m_c+m_u}{m_B}\left(-3.2445 + 3.3231 \frac{m_B}{m_0}  \right. \nonumber\\
&   \left. - 1.0442 \left(\frac{m_B}{m_0}\right)^2+ 0.1333 \left(\frac{m_B}{m_0}\right)^3 \right.\nonumber\\
&\left. -  0.005737 \left(\frac{m_B}{m_0}\right)^4\right),\nonumber\\
{c}_{fl_2}=&\left(\frac{m_B}{m_c+m_u}\right)^{0.3}\left(1.4697- 0.3188 \frac{m_B}{m_0} \right. \nonumber\\
&   \left. +  0.04506 \left(\frac{m_B}{m_0}\right)^2 - 0.002205 \left(\frac{m_B}{m_0}\right)^3\right),\nonumber\\
{c}_{0l_2}=&\frac{m_c+m_u}{m_B}\left(-5.1878 + 5.2976 \frac{m_B}{m_0} \right. \nonumber\\
&   \left. - 1.6633 \left(\frac{m_B}{m_0}\right)^2 + 0.2127 \left(\frac{m_B}{m_0}\right)^3 \right. \nonumber\\
&   \left. - 0.009171 \left(\frac{m_B}{m_0}\right)^4\right).
\end{align}
Here, $m_B=m_u+m_{H}$. $m_H$ represents the mass of diquark for doubly heavy triquark $(\bar{u}(cc))$ or $(\bar{u}(bb))$.

\begin{table}[!phtb]
\caption{Parameter values \cite{Feng:2023txx,Faustov:2021hjs}.}  \label{tab:parmv}
\centering
\begin{tabular*}{0.47\textwidth}{@{\extracolsep{\fill}}cl@{}}
\hline\hline
          & $m_{u}=0.33\; {\gev}$, \; $m_b=4.88\; {\gev}$  \\
          & $m_c=1.55\; {\gev}$, \; $\sigma=0.18\; {\gev^2}$,\; $C=-0.3\; {\gev}$ \\
          & $c_{fn_{r}}=1$,\; $c_{fl}=1.17$\\
$(bb)$    & $c_{0n_{r}}(1^3s_1)=0.01$,\;  $c_{0n_{l}}(1^3s_1)=0.001$\\
$(cc)$    & $c_{0n_{r}}(1^3s_1)=0.205$, \; $c_{0n_{l}}(1^3s_1)=0.337$\\
\hline
\hline
\end{tabular*}
\end{table}

Determining the {\rts} for the quadruply heavy pentaquark $bb\bar{u}cc$ requires not only the fundamental parameters $m_u$, $m_b$, $m_c$, $\sigma$ and $C$, but also the {\rt} parameters $c_{fx}$ and $c_{0x}$ for diquarks, triquarks, and pentaquarks. The fundamental parameters listed in Table \ref{tab:parmv} and adopted from Refs. \cite{Feng:2023txx,Faustov:2021hjs} are universal for all bound states. These parameters are also adopted to fit the relations in Eqs. (\ref{fitcfxnr}) and (\ref{apprelA1}). The {\rt} parameters in Table \ref{tab:parmv} are obtained by fitting {\rts} for doubly heavy mesons and are further utilized to determine the {\rts} for doubly heavy diquarks \cite{Feng:2023txx}. The relations in Eq. (\ref{fitcfxnr}) are obtained by fitting the {\rts} for doubly heavy mesons, triply heavy baryons and doubly heavy tetraquarks \cite{Xie:2024dfe}. These relations are employed to calculate the {\rt} parameters $c_{fN_{r}}$, $c_{0N_{r}}$, $c_{fL}$, and $c_{0L}$ for bound states composed of two heavy clusters. The relations in Eq. (\ref{apprelA1}) are obtained by fitting the {\rts} for heavy-light mesons and doubly heavy baryons \cite{Liu:2026npi}, and they serve to determine the {\rt} parameters for bound states composed of one heavy cluster and one light cluster. The treatment of these parameters and relations is logically self-consistent. These parameters and relations have been successfully applied to investigate the {\rts} for mesons, diquarks, baryons, triquarks, tetraquarks, pentaquarks, and hexaquarks \cite{Chen:2022flh,Xie:2024lfo,Feng:2023txx,Xie:2024dfe,Song:2025cla,Liu:2026npi,
Chen:2025fyh,Song:2024bkj}. Herein, they are employed to study the quadruply heavy pentaquark $bb\bar{u}cc$, yielding results consistent with other theoretical predictions (see the discussions in Sec. \ref{subsec:disc}). This validates the universality of the present Regge trajectory relations and parameter values.

\subsection{$\lambda$-{\trs} for the {\ptq}}\label{subsec:rts}

\begin{table}[!phtb]
\caption{Spin-averaged masses (in {\gev}) of radially and orbitally $\lambda$-excited {\ptq}. Configurations $(bb)(\bar{u}(cc))$ and $(cc)(\bar{u}(bb))$ are considered. $|n_1^{2s_1+1}l_{1j_1},n_2l_{2},n_3^{2s_{3}+1}l_{3j_3},NL\rangle$ denotes the spin-averaged states. All masses are calculated via Eqs. (\ref{ppt2q}), (\ref{pppa2qQ}), (\ref{fitcfxnr}), and (\ref{apprelA1}).}  \label{tab:lambda}
\centering
\begin{tabular*}{0.450\textwidth}{@{\extracolsep{\fill}}ccc@{}}
\hline\hline
  $|n_1^{2s_1+1}l_{1j_1},n_2l_{2},n_3^{2s_{3}+1}l_{3j_3},NL\rangle$       & $(bb)(\bar{u}(cc))$   &  $(cc)(\bar{u}(bb))$  \\
\hline
 $|1^3s_1, 1s, 1^3s_1, 1S\rangle$  &13.11    &13.15    \\
 $|1^3s_1, 1s, 1^3s_1, 2S\rangle$  &13.56    &13.61    \\
 $|1^3s_1, 1s, 1^3s_1, 3S\rangle$  &13.86    &13.92    \\
 $|1^3s_1, 1s, 1^3s_1, 4S\rangle$  &14.12    &14.18    \\
 $|1^3s_1, 1s, 1^3s_1, 5S\rangle$  &14.35    &14.42    \\
$|1^3s_1, 1s, 1^3s_1, 1S\rangle$  &13.11    &13.16    \\
$|1^3s_1, 1s, 1^3s_1, 1P\rangle$  &13.43    &13.48    \\
$|1^3s_1, 1s, 1^3s_1, 1D\rangle$  &13.66    &13.71    \\
$|1^3s_1, 1s, 1^3s_1, 1F\rangle$  &13.85    &13.91    \\
$|1^3s_1, 1s, 1^3s_1, 1G\rangle$  &14.02    &14.08    \\
\hline\hline
\end{tabular*}
\end{table}

\begin{table*}[!phtb]
\caption{Fitted formulas for the $\lambda$-, $\rho$-, and $\sigma$-{\trs} of the {\ptq} in the $(bb)(\bar{u}(cc))$ and $(cc)(\bar{u}(bb))$ configurations. ${\ast}$ indicates that alternative fitted formulas are better than the listed ones.}  \label{tab:formulas}
\centering
\begin{tabular*}{1.0\textwidth}{@{\extracolsep{\fill}}cll@{}}
\hline\hline
        & $(bb)(\bar{u}(cc))$  &  $(cc)(\bar{u}(bb))$   \\
\hline
 $\lambda$  & $M=12.9934+ 0.5286 (0.1029+ N_r)^{2/3}$
                    & $M=13.0146+ 0.5465 (0.12562+ N_r)^{2/3}$        \\
            & $M=12.9934+ 0.3973 (0.1655+ L)^{2/3}$
                    &$M=13.0146+ 0.4103 (0.2032+ L)^{2/3}$         \\
 $\rho_1$   & $M=13.0941+ 0.3309 (0.0099+ n_{r_1})^{2/3}$
                    & $M=12.9835+ 0.4683 (0.2169+ n_{r_1})^{2/3}$        \\
            & $M=13.0941+ 0.2603 (0.0009+ l_1)^{2/3}$
                    &$M=12.9843+ 0.3682 (0.3519+ l_1)^{2/3}$         \\
 $\rho_2$   &$M=12.7518+ 0.7144 \sqrt{0.2509+ n_{r_2}}$
                    &$M=12.7633+ 0.7248 \sqrt{0.2879+ n_{r_2}}$         \\
            &$M=12.7522+ 0.5654 \sqrt{0.4019+ l_2}$
                    &$M=12.7637+ 0.5776 \sqrt{0.4699+ l_2}$         \\
 $\sigma$   &$M=12.923+ 0.4411 (0.2829+ n_{r_3})^{2/3}$
                   &$M=13.1627+ 0.2612 (0.0009+ n_{r_3})^{2/3}$$^{\ast}$         \\
            &$M=12.9401+ 0.3428 (0.3969+ l_3)^{2/3}$
                   &$M=13.1498+ 0.2132 (0.0009+ l_3)^{2/3}$$^{\ast}$         \\
\hline\hline
\end{tabular*}
\end{table*}

\begin{figure*}[!phtb]
\centering
\subfigure[]{\label{subfigure:cfa}\includegraphics[scale=0.44]{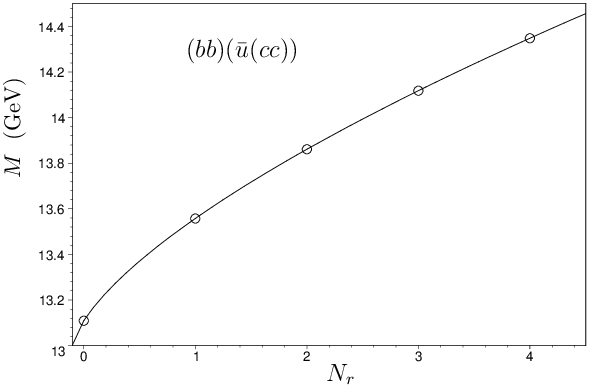}}
\subfigure[]{\label{subfigure:cfa}\includegraphics[scale=0.44]{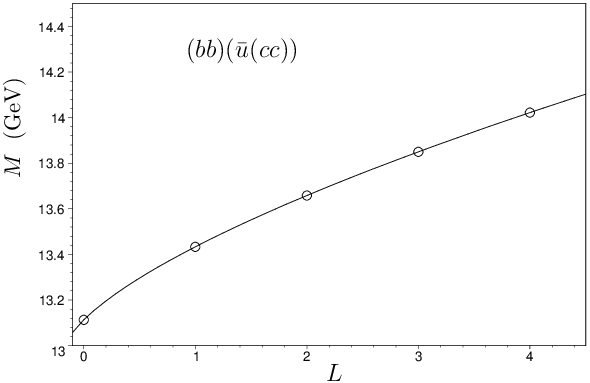}}
\subfigure[]{\label{subfigure:cfa}\includegraphics[scale=0.44]{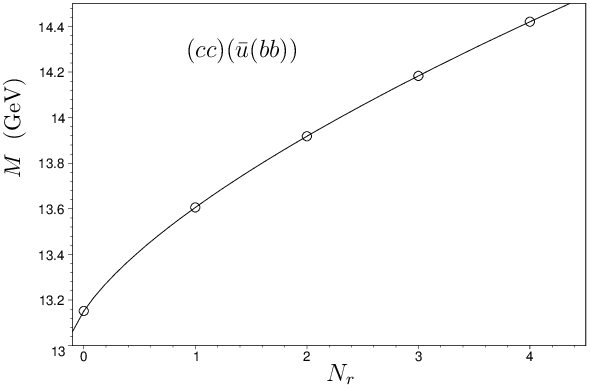}}
\subfigure[]{\label{subfigure:cfa}\includegraphics[scale=0.44]{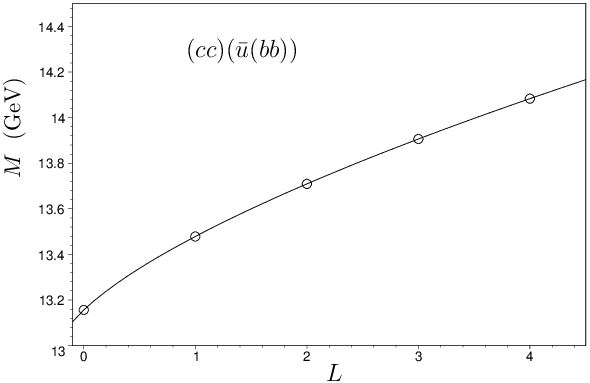}}
\caption{$\lambda$-{\trs} for the {\ptq} in $(bb)(\bar{u}(cc))$ and $(cc)(\bar{u}(bb))$ configurations. (a) and (b) are for $(bb)(\bar{u}(cc))$ configuration; (c) and (d) are for $(cc)(\bar{u}(bb))$ configuration. $N_{r}$ and ${L}$ are the radial and orbital quantum numbers for the $\lambda$-mode, respectively. Circles represent the predicted data listed in Table \ref{tab:lambda}, evaluated via either Eqs. (\ref{ppt2q}) and (\ref{pppa2qQ}) or Eqs. (\ref{rtpf}) and (\ref{pppa2qQ}). The black lines correspond to the fitted formulas, obtained by fitting the calculated data in Table \ref{tab:lambda}; these formulas are summarized in Table \ref{tab:formulas}. }\label{fig:lambda}
\end{figure*}

When calculating the $\lambda$-mode, all other modes are kept in their radially ground states, and the parameters used correspond to the radial ground states of those modes. According to Eqs. (\ref{pppa2qQ}) and (\ref{rtpf}), we can see that terms corresponding to $\rho_1$-, $\rho_2$-, and $\sigma$-{\trs} are constant when considering $\lambda$-{\trs}; therefore, Eq. (\ref{rtpf}) takes the simplest form. The $\lambda$-{\trs} are the simplest among four series of {\rts} for the {\ptq}. The {\rt} behavior can read directly from Eq. (\ref{rtpf}), $M{\sim}x^{2/3}_{\lambda}$, where $x_{\lambda}=N_r,\,L$.

We compute the spin-averaged masses of radially and orbitally $\lambda$-excited states for both configurations $(bb)(\bar{u}(cc))$ and $(cc)(\bar{u}(bb))$ using Eqs. (\ref{rtpf}), (\ref{pppa2qQ}), (\ref{fitcfxnr}), (\ref{apprelA1}), and parameter values listed in Table \ref{tab:parmv}. The calculated masses are listed in Table \ref{tab:lambda} and the {\rt} formulas are given in Table \ref{tab:formulas}.
For heavy-heavy systems, the binding energy decreases with the reduced mass. The reduced mass $\mu_{\lambda}$ of the configuration $(cc)(\bar{u}(bb))$ is smaller than that of $(bb)(\bar{u}(cc))$; accordingly, the masses of the $(cc)(\bar{u}(bb))$ configuration are slightly larger than those of the $(bb)(\bar{u}(cc))$ configuration (see $\beta_{L}$ and $\beta_{N_r}$ in Eq. (\ref{pppa2qQ})). For the ground state, the {\rt} parameters $\beta_{\lambda}$, $\beta_{\rho_1}$, $\beta_{\rho_2}$, $\beta_{\sigma}$, $c_{0x_{\lambda}}$, $c_{0x_{\rho_1}}$, $c_{0x_{\rho_2}}$, and $c_{0x_{\sigma}}$ are nonzero, even though the corresponding quantum numbers $x_{\lambda}$, $x_{\rho_1}$, $x_{\rho_2}$, $x_{\sigma}$ are zero. The {\rt} parameters (see Eqs. (\ref{pppa2qQ}), (\ref{fitcfxnr}), and (\ref{apprelA1})) differ between the $(cc)(\bar{u}(bb))$ and $(bb)(\bar{u}(cc))$ configurations. Consequently, these two configurations possess different ground state masses (see Eq. (\ref{rtpf})).

Figure \ref{fig:lambda} shows the radial and orbital $\lambda$-{\trs} of the {\ptq} for the configurations $(bb)(\bar{u}(cc))$ and $(cc)(\bar{u}(bb))$. Circles represent the predicted data calculated via the complete forms of the {\rt} relations in Eqs. (\ref{rtpf}) and (\ref{pppa2qQ}). The black lines represent the fitted formulas, which are obtained by fitting the calculated data in Table \ref{tab:lambda}. For the $\lambda$-mode, the fitted formulas are identical to the complete forms, which are listed in Table \ref{tab:formulas}.

\subsection{$\rho$-{\trs} for the {\ptq}}\label{subsec:rts}

\begin{table*}[!phtb]
\caption{Same as Table \ref{tab:lambda} except for the $\rho_1$- and $\rho_2$-excited states. States marked with $(\times)$ do not exist.}  \label{tab:rho}
\centering
\begin{tabular*}{1.0\textwidth}{@{\extracolsep{\fill}}lccccc@{}}
\hline\hline
$|n_1^{2s_1+1}l_{1j_1},n_2l_{2},n_3^{2s_{3}+1}l_{3j_3},NL\rangle$       & $(bb)(\bar{u}(cc))$   &  $(cc)(\bar{u}(bb))$
&$|n_1^{2s_1+1}l_{1j_1},n_2l_{2},n_3^{2s_{3}+1}l_{3j_3},NL\rangle$       & $(bb)(\bar{u}(cc))$   &  $(cc)(\bar{u}(bb))$  \\
\hline
 $|1^3s_1, 1s, 1^3s_1, 1S\rangle$  &13.11    &13.15
               & $|1^3s_1, 1s, 1^3s_1, 1S\rangle$  &13.11    &13.15   \\
 $|2^3s_1, 1s, 1^3s_1, 1S\rangle$  &13.43    &13.52
               & $|1^3s_1, 2s, 1^3s_1, 1S\rangle$  &13.55    &13.59   \\
 $|3^3s_1, 1s, 1^3s_1, 1S\rangle$  &13.62    &13.78
               & $|1^3s_1, 3s, 1^3s_1, 1S\rangle$  &13.82    &13.86   \\
 $|4^3s_1, 1s, 1^3s_1, 1S\rangle$  &13.78    &14.00
               & $|1^3s_1, 4s, 1^3s_1, 1S\rangle$  &14.04    &14.08   \\
 $|5^3s_1, 1s, 1^3s_1, 1S\rangle$  &13.93    &14.21
               & $|1^3s_1, 5s, 1^3s_1, 1S\rangle$  &14.22    &14.26   \\
\hline
 $|1^3s_1, 1s, 1^3s_1, 1S\rangle$  &13.10    &13.17
             & $|1^3s_1, 1s, 1^3s_1, 1S\rangle$  &13.11    &13.16 \\
 $|1^3p_2, 1s, 1^3s_1, 1S\rangle(\times)$  &13.35    &13.43
             & $|1^3s_1, 1p, 1^3s_1, 1S\rangle$  &13.42    &13.46 \\
 $|1^3d_3, 1s, 1^3s_1, 1S\rangle$  &13.51    &13.64
             & $|1^3s_1, 1d, 1^3s_1, 1S\rangle$  &13.63    &13.67 \\
 $|1^3f_4, 1s, 1^3s_1, 1S\rangle(\times)$  &13.64    &13.81
             &  $|1^3s_1, 1f, 1^3s_1, 1S\rangle$  &13.80    &13.84 \\
 $|1^3g_5, 1s, 1^3s_1, 1S\rangle$  &13.75    &13.97
             & $|1^3s_1, 1g, 1^3s_1, 1S\rangle$  &13.94    &13.98 \\
\hline\hline
\end{tabular*}
\end{table*}

\begin{figure*}[!phtb]
\centering
\subfigure[]{\label{subfigure:cfa}\includegraphics[scale=0.44]{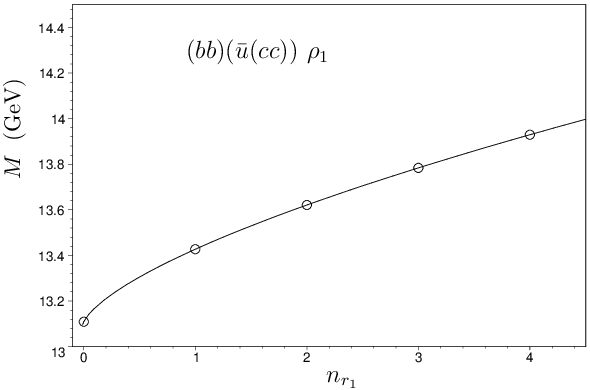}}
\subfigure[]{\label{subfigure:cfa}\includegraphics[scale=0.44]{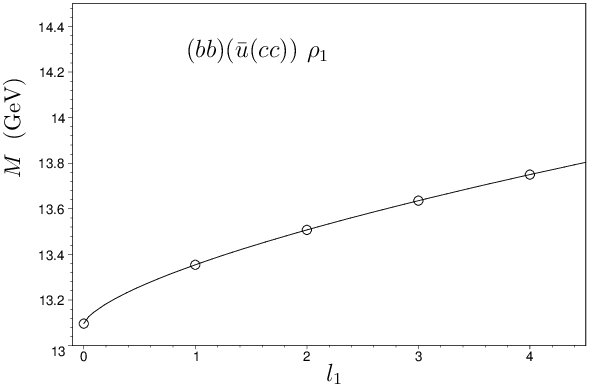}}
\subfigure[]{\label{subfigure:cfa}\includegraphics[scale=0.44]{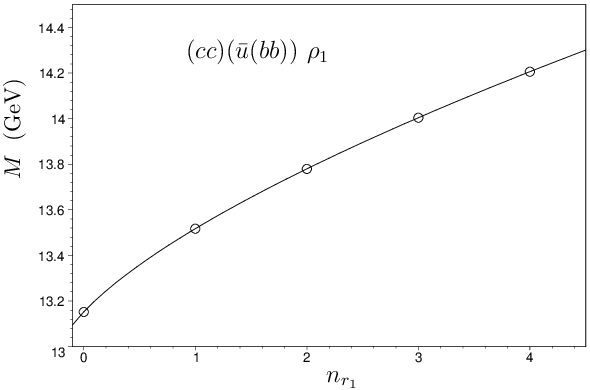}}
\subfigure[]{\label{subfigure:cfa}\includegraphics[scale=0.44]{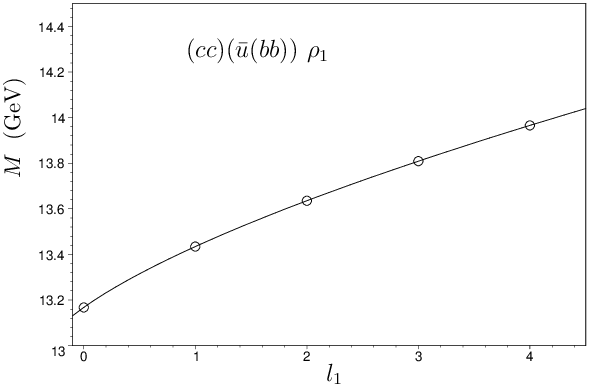}}
\subfigure[]{\label{subfigure:cfa}\includegraphics[scale=0.44]{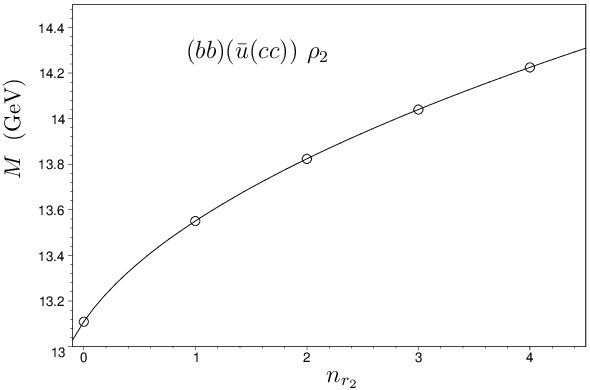}}
\subfigure[]{\label{subfigure:cfa}\includegraphics[scale=0.44]{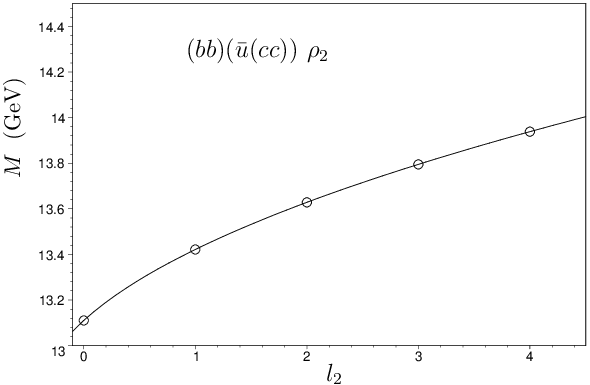}}
\subfigure[]{\label{subfigure:cfa}\includegraphics[scale=0.44]{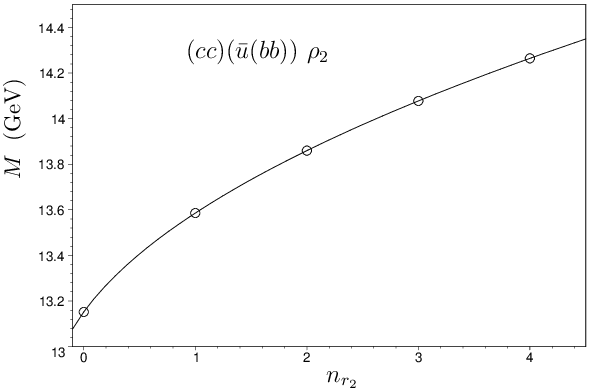}}
\subfigure[]{\label{subfigure:cfa}\includegraphics[scale=0.44]{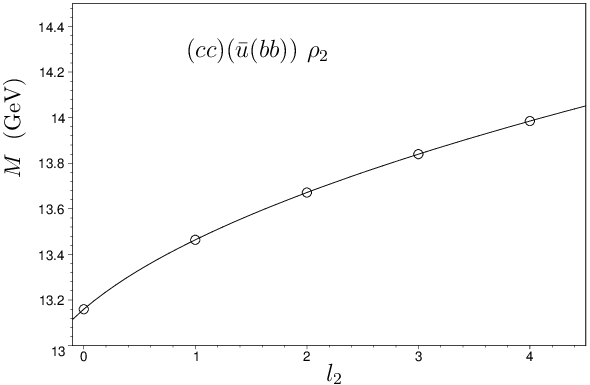}}
\caption{$\rho_1$- and $\rho_2$-{\trs} for the {\ptq} in the $(bb)(\bar{u}(cc))$ and $(cc)(\bar{u}(bb))$ configurations. $\rho_1$ trajectories are in (a-d); $\rho_2$ trajectories are in (e-h). $n_{r_1}$ and ${l_1}$ stand for the radial and orbital quantum numbers of the $\rho_1$-mode, respectively, while $n_{r_2}$ and ${l_2}$ are the corresponding numbers for the $\rho_2$-mode.  Circles represent the predicted data listed in Table \ref{tab:rho}, calculated using either Eqs. (\ref{ppt2q}) and (\ref{pppa2qQ}) or Eqs. (\ref{rtpf}) and (\ref{pppa2qQ}). The black lines correspond to the fitted formulas, obtained by fitting the predicted data in Table \ref{tab:rho}; these formulas are summarized in Table \ref{tab:formulas}. }\label{fig:rho}
\end{figure*}

The $\rho_1$- and $\rho_2$-{\tr} relations are obtained by using Eqs. (\ref{rtpf}), (\ref{pppa2qQ}), (\ref{fitcfxnr}), (\ref{apprelA1}), and parameter values listed in Table \ref{tab:parmv}. When considering the $\rho_1$-{\trs}, all remaining modes are fixed at their radially ground states, and the adopted parameters correspond to these radial ground states. By a similar procedure, the $\rho_2$-{\trs} can be determined. The $\rho$-{\trs} are shown in Fig. \ref{fig:rho}. Spin-averaged masses of radially and orbitally $\rho_1$- and $\rho_2$-excited states can be calculated by using the $\rho_1$- and $\rho_2$-{\tr} relations. The calculated results are listed in Table \ref{tab:rho}. In Table \ref{tab:rho}, entries labeled with $(\times)$ represent nonexist states: the $1^3p_2$ and $1^3f_4$ diquark states of $(bb)$ and $(cc)$ are forbidden by the Pauli principle \cite{Feng:2023txx}.
The $\rho_1$-excited masses of the $(cc)(\bar{u}(bb))$ and $(bb)(\bar{u}(cc))$ configurations increase with $n_{r_1}$ and $l_1$. The former grow more rapidly than the latter, as the masses of the excited diquark $(cc)$ increase faster than those of excited diquark $(bb)$. For $\rho_2$-excited states, the masses of the $(cc)(\bar{u}(bb))$ configuration are slightly larger than those of the $(bb)(\bar{u}(cc))$ configuration, since the reduced mass $\mu_{\lambda}$ of $(cc)(\bar{u}(bb))$ is smaller than that of $(bb)(\bar{u}(cc))$.

From Eqs. (\ref{rtpf}), (\ref{pppa2qQ}), (\ref{fitcfxnr}), and (\ref{apprelA1}), we can see that for $\rho_1$-{\trs}, terms corresponding to the $\rho_2$- and $\sigma$-modes are constants. In contrast, term corresponding to $\lambda$-mode depends on $x_{\rho_1}$ via $\beta_{\lambda}$ and $c_{0x_{\lambda}}$ even though $x_{\lambda}$ itself is a constant. This makes the $\rho_1$-{\tr} expressions far longer and more cumbersome than those for $\lambda$-{\trs}.
As an illustration, the complete form of the radial $\rho_1$-{\tr} for the configuration $(bb)(\bar{u}(cc))$ reads
\begin{widetext}
\begin{align}\label{fullr}
M=&12.9779+ 0.332799 (0.01+ n_{r_1})^{2/3} +
0.461167\left(\frac{ 9.61+ 0.332799 (0.01+ n_{r_1})^{2/3}}
           { 13.2779+ 0.332799 (0.01+ n_{r_1})^{2/3}}\right)^{-1/3} \nonumber\\
  &\times\left[0.334- \frac{ 0.31911 \left(9.61+ 0.332799 (0.01+ n_{r_1})^{2/3}\right)}
                   { 13.2779+ 0.332799 (0.01+ n_{r_1})^{2/3}}\right]^{2/3}
        \left[1.008+ \frac{ 0.0293435 \left(9.61+ 0.332799 (0.01+ n_{r_1})^{2/3}\right)}
        {13.2779+  0.332799 (0.01+ n_{r_1})^{2/3}}\right].
\end{align}
\end{widetext}
Fitting the calculated results with Eq. (\ref{fullr}) yields the simple form
\bea\label{fitrhoa}
M=13.0941+ 0.3309 (0.0099+ n_{r_1})^{2/3}.
\eea
The radial {\rt} for diquark $(bb)$ can be obtained from Eqs. (\ref{ppt2q}) and (\ref{pppa2qQ}):
\bea\label{rtdiq}
M=9.61+0.332799 (0.01+ n_{r_1} )^{2/3}.
\eea
Equation (\ref{fullr}) is tedious and the behavior respective to $n_{r_1}$ is not obvious whereas the dependence of (\ref{fitrhoa}) on $n_{r_1}$ is evident. Since Eq. (\ref{fullr}) can be well approximated by (\ref{fitrhoa}), both expressions share identical behavior respective to $n_{r_1}$, namely $M{\sim}n_{r_1}^{2/3}$.
Equation (\ref{rtdiq}) is the radial {\rt} relation for diquark $(bb)$. Clearly, the diquark $(bb)$ {\rt} differ from the $\rho_1$-{\tr} given in Eq. (\ref{fullr}), yet they have the same {\rt} behaviors.

\begin{table}[!phtb]
\caption{Same as Table \ref{tab:lambda} except for the $\sigma$-excited states. States marked with $(\times)$ do not exist.}  \label{tab:sigma}
\centering
\begin{tabular*}{0.45\textwidth}{@{\extracolsep{\fill}}lcc@{}}
\hline\hline
  $|n_1^{2s_1+1}l_{1j_1},n_2l_{2},n_3^{2s_{3}+1}l_{3j_3},NL\rangle$       & $(bb)(\bar{u}(cc))$   &  $(cc)(\bar{u}(bb))$  \\
\hline
 $|1^3s_1, 1s, 1^3s_1, 1S\rangle$  &13.11    &13.15    \\
 $|1^3s_1, 1s, 2^3s_1, 1S\rangle$  &13.45    &13.43    \\
 $|1^3s_1, 1s, 3^3s_1, 1S\rangle$  &13.69    &13.58    \\
 $|1^3s_1, 1s, 4^3s_1, 1S\rangle$  &13.90    &13.71    \\
 $|1^3s_1, 1s, 5^3s_1, 1S\rangle$  &14.09    &13.82    \\
 $|1^3s_1, 1s, 1^3s_1, 1S\rangle$  &13.12    &13.14    \\
 $|1^3s_1, 1s, 1^3p_2, 1S\rangle(\times)$  &13.37    &13.36    \\
 $|1^3s_1, 1s, 1^3d_3, 1S\rangle$  &13.55    &13.49    \\
 $|1^3s_1, 1s, 1^3f_4, 1S\rangle(\times)$  &13.71    &13.60    \\
 $|1^3s_1, 1s, 1^3g_5, 1S\rangle$  &13.86    &13.69    \\
\hline\hline
\end{tabular*}
\end{table}

\begin{figure*}[!phtb]
\centering
\subfigure[]{\label{subfigure:cfa}\includegraphics[scale=0.44]{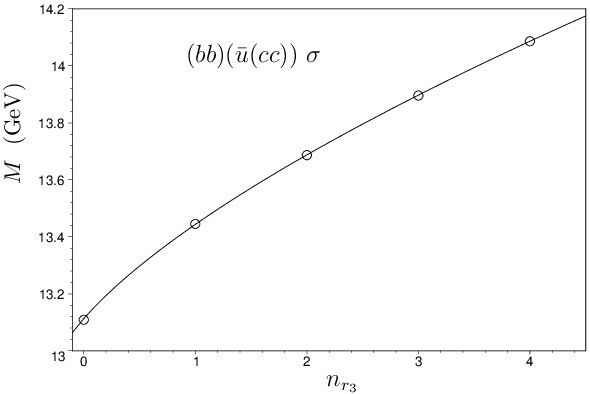}}
\subfigure[]{\label{subfigure:cfa}\includegraphics[scale=0.44]{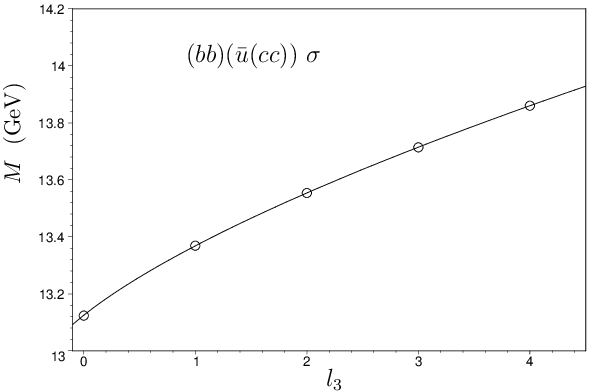}}
\subfigure[]{\label{subfigure:cfa}\includegraphics[scale=0.44]{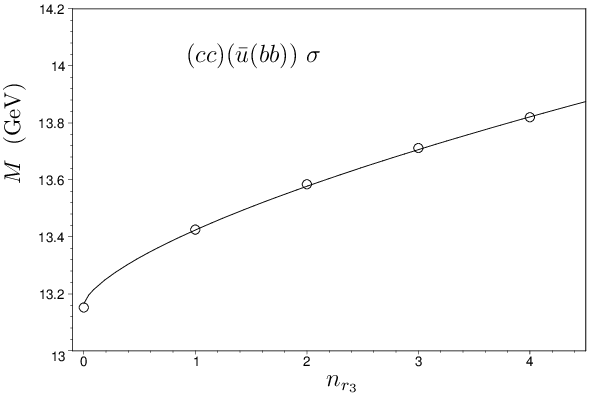}}
\subfigure[]{\label{subfigure:cfa}\includegraphics[scale=0.44]{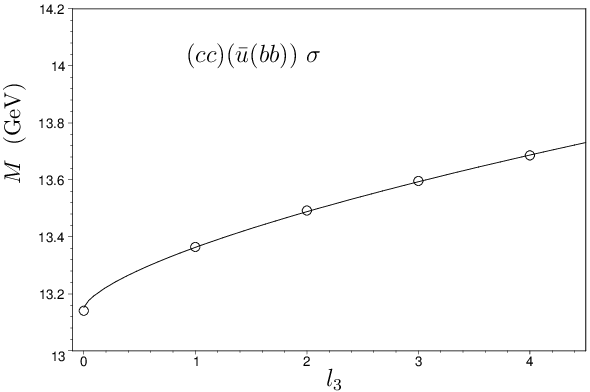}}
\caption{The $\sigma$-{\trs} for the {\ptq} in configurations $(bb)(\bar{u}(cc))$ and $(cc)(\bar{u}(bb))$. (a) and (b) are for $(bb)(\bar{u}(cc))$ configuration; (c) and (d) are for $(cc)(\bar{u}(bb))$ configuration. $n_{r_3}$ and ${l_3}$ are the radial and orbital quantum numbers for the $\sigma$-mode, respectively. Circles represent the predicted data listed in Table \ref{tab:sigma}, obtained from Eqs. (\ref{ppt2q}) and (\ref{pppa2qQ}) or from Eqs. (\ref{rtpf}) and (\ref{pppa2qQ}). The black dashed lines correspond to the fitted formulas, obtained by fitting the calculated data in Table \ref{tab:sigma}; these formulas are listed in Table \ref{tab:formulas}. }\label{fig:sigma}
\end{figure*}

\begin{table}[!phtb]
\caption{Comparison between two fitted formulas. The masses of orbitally $\sigma$-excited states in the $(cc)(\bar{u}(bb))$ configuration are calculated using the complete form of {\rts} in Eqs. (\ref{pppa2qQ}) and (\ref{rtpf}) (CF). By fitting the calculated masses, we obtain the fitted formulas $M=13.1498+ 0.213174 (0.0009+ l_3)^{2/3}$ (A) and $M=13.0979+ 0.25979 (0.0519+ l_3)^{7/12}$ (B).}  \label{tab:sigmacomp}
\centering
\begin{tabular*}{0.49\textwidth}{@{\extracolsep{\fill}}lcccc@{}}
\hline\hline
  $|n_1^{2s_1+1}l_{1j_1},n_2l_{2},n_3^{2s_{3}+1}l_{3j_3},NL\rangle$         &  CF & A & B \\
\hline
 $|1^3s_1, 1s, 1^3s_1, 1S\rangle$              &13.1408  &13.1518    &13.1441  \\
 $|1^3s_1, 1s, 1^3p_2, 1S\rangle(\times)$      &13.3646  &13.3631    &13.3655   \\
 $|1^3s_1, 1s, 1^3d_3, 1S\rangle$              &13.4923  &13.4883    &13.4930   \\
 $|1^3s_1, 1s, 1^3f_4, 1S\rangle(\times)$      &13.5962  &13.5933    &13.5959   \\
 $|1^3s_1, 1s, 1^3g_5, 1S\rangle$              &13.6860  &13.6871    &13.6855   \\
\hline\hline
\end{tabular*}
\end{table}

\begin{figure}[!phtb]
\centering
\includegraphics[width=0.35\textheight]{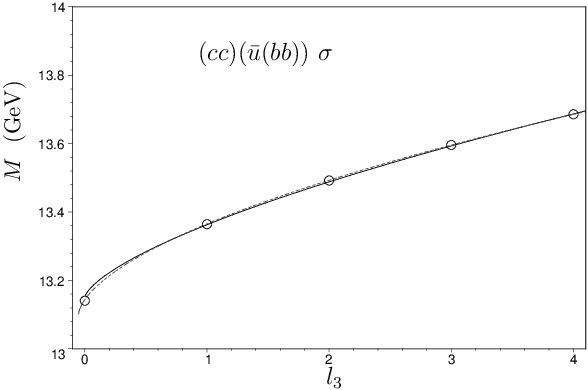}
\caption{Comparison of two approximation results. Circles represent data calculated by using the complete form of the orbital $\sigma$-{\tr}, which are listed in Table \ref{tab:sigmacomp}. The solid line represents the fitted formula $M=13.1498+ 0.213174 (0.0009+ l_3)^{2/3}$. The dashed line represents the fitted formula $M=13.0979+ 0.25979 (0.0519+ l_3)^{7/12}$.   }\label{fig:compf}
\end{figure}

\subsection{$\sigma$-{\trs} for the {\ptq}}
Among four series of {\rts} for the {\ptq}, $\sigma$-{\trs} are the most tedious. Following a procedure analogous to that used for $\lambda$- and $\rho$-{\trs}, we derive the $\sigma$-{\trs} using Eqs. (\ref{rtpf}), (\ref{pppa2qQ}), (\ref{fitcfxnr}), and (\ref{apprelA1}) together with the parameter values listed in Table \ref{tab:parmv}. The $\sigma$-{\trs} are shown Fig. \ref{fig:sigma} and the calculated masses are summarized in Table \ref{tab:sigma}.
The $\sigma$-excited masses of the $(cc)(\bar{u}(bb))$ and $(bb)(\bar{u}(cc))$ configurations increase with $n_{r_3}$ and $l_3$. The former grow more rapidly than the latter, as the masses of the excited diquark $(cc)$ increase faster than those of the excited diquark $(bb)$.

The cumbersome $\rho_1$-, $\rho_2$-, and $\sigma$-{\trs} can be approximated by simple functions. We prefer to adopt the functional form given in Eq. (\ref{rtpf}) for function approximation, even though other functional forms can sometimes yield decent or even superior approximation results.
For the $(bb)(\bar{u}(cc))$ configuration, the complete form of radial and orbital $\sigma$-{\trs} can be well approximated by the simple formulas which take the functional form in Eq. (\ref{rtpf}). In contrast, for the $(cc)(\bar{u}(bb))$ configuration, $M=12.9774+ 0.41051 \sqrt{0.1979+ n_{r_3}}$ provides a better fit to the complete form than $M=13.1627+ 0.26124 (0.0009+ n_{r_3})^{2/3}$. Likewise, $M=13.0979+ 0.25979 (0.0519+ l_3)^{7/12}$ is better than $M=13.1498+ 0.213174 (0.0009+ l_3)^{2/3}$. Table \ref{tab:sigmacomp} and Fig. \ref{fig:compf} show comparison between the two sets of fitting results.

Observing Eq. (\ref{rtpf}), we find that for the $\rho$- and $\sigma$-{\trs}, the corresponding diquark or triquark {\rts} govern the behavior associated with orbital and radial quantum numbers when contributions from other functional components are negligible. In this case, fitting functions constructed following the functional form of these diquark/triquark {\rts} provide a good approximation of the full expression. Conversely, when contributions from other functional components are large, fitting functions with functional forms distinct from the relevant diquark or triquark {\rts} yield superior results (see Table \ref{tab:sigmacomp} and Fig. \ref{fig:compf}).
Even for this case, we still prefer to adopt the functional form of the corresponding diquark/triquark {\rts} because the differences between different functional form are small, see Table \ref{tab:sigmacomp} and Fig. \ref{fig:compf}.
In Table \ref{tab:formulas}, the listed $\sigma$-{\trs} for $(cc)(\bar{u}(bb))$ configuration take the functional form of the corresponding diquark {\rts}.

\subsection{Discussions}\label{subsec:disc}

It is worth emphasizing that various functional forms can deliver good fitting results given a limited fitting range and a finite number of data points (see , e.g., Tables 1 and 2 in Ref. \cite{Chen:2022flh}). For this reason, the functional behavior of Regge trajectories for diquarks and triquark offers guidance for fitting the pentaquark. This conclusion can be generalized to other bound-state systems. The Regge trajectories of internal constituents can provide guidance for the distinct series of Regge trajectories corresponding to the entire bound state \cite{Song:2025cla,Chen:2025fyh,Liu:2026npi,
Xie:2024lfo,Song:2024bkj,Xie:2024dfe,Liu:2026vpi}.

In Ref. \cite{An:2021vwi}, the masses of pentaquark $bbcc\bar{n}$ are calculated using the Chromomagnetic interaction model. The spin-averaged mass of the ground state is $13.24$ {\gev}. Our results are consistent with theoretical predictions in Ref. \cite{An:2021vwi}. The spin-averaged ground state mass is $13.11$ {\gev} in the $(bb)(\bar{u}(cc))$ configuration and $13.15$ {\gev} in the $(cc)(\bar{u}(bb))$ configuration.

\section{Conclusions}\label{sec:conc}
Systematic investigations of four series of {\rts} for {\QH} pentaquarks are still lacking. Using the diquark and triquark {\rt} relations, we propose the Regge trajectory relations for the quadruply heavy pentaquark ${bb\bar{u}cc}$: $M=2m_{b}+2m_{c}+m_{u}+5C/2 +\beta_{x_{\lambda}}(x_{\lambda}+c_{0x_{\lambda}})^{2/3}
+\beta_{x_{\rho_1}}(x_{\rho_1}+c_{0x_{\rho_1}})^{2/3}+\beta_{x_{\rho_2}}\sqrt{x_{\rho_2}+c_{0x_{\rho_2}}}
+\beta_{x_{\sigma}}(x_{\sigma}+c_{0x_{\sigma}})^{2/3}$. Four series of Regge trajectories, namely, the $\lambda$-, $\rho_1$-, $\rho_2$-, and $\sigma$-trajectories, are investigated.
Using the proposed pentaquark {\rt} relations, we provide rough estimates for spin-averaged masses of the $\lambda$-, $\rho_1$-, $\rho_2$-, and $\sigma$-excited states.

We demonstrate that accounting for the internal structure and substructure of pentaquarks is indispensable for constructing the $\rho_1$-, $\rho_2$-, and $\sigma$-trajectories; without such structural considerations, functional forms and trajectory parameters can only be obtained via pure fitting against theoretical or experimental data.
We further prove that the Regge trajectories of diquark 1, triquark, and diquark 2 (embedded within the triquark) do not correspond one-to-one to the $\rho_1$-, $\rho_2$-, and $\sigma$-trajectories. Nevertheless, these trajectories govern the behaviors of the respective $\rho_1$-, $\rho_2$-, and $\sigma$-trajectories.
For both configurations $(bb)(\bar{u}(cc))$ and $(cc)(\bar{u}(bb))$, the $\lambda$-, $\rho_1$-, and $\sigma$-trajectories exhibit behavior of $M{\sim}x^{2/3}$ ($x=n_{r_1},n_{r_3},l_1,l_3,N_{r},L$), whereas the $\rho_2$-trajectories exhibit behavior of $M{\sim}\sqrt{x}$ ($x=n_{r_2},\,l_2$).
The functional behavior of Regge trajectories for diquarks and triquark offers guidance for fitting the pentaquark {\rts}.

\section*{DATA AVAILABILITY}
There are no publicly available research data or software supporting this manuscript. Requests for further information or data should be sent to the authors.






\begin{thebibliography}{99}

\bibitem{takahashi:pdg2026}
F. Takahashi et al. (Particle Data Group), Int. J. Mod. Phys. A 41 , 2630011 (2026)

\bibitem{Brambilla:2019esw}
N.~Brambilla, S.~Eidelman, C.~Hanhart, A.~Nefediev, C.~P.~Shen, C.~E.~Thomas, A.~Vairo and C.~Z.~Yuan,
Phys. Rept. \textbf{873}, 1-154 (2020)
doi:10.1016/j.physrep.2020.05.001
[arXiv:1907.07583 [hep-ex]].

\bibitem{Jaffe:2004ph}
R.~L.~Jaffe,
Phys. Rept. \textbf{409}, 1-45 (2005)
doi:10.1016/j.physrep.2004.11.005
[arXiv:hep-ph/0409065 [hep-ph]].

\bibitem{Liu:2019zoy}
Y.~R.~Liu, H.~X.~Chen, W.~Chen, X.~Liu and S.~L.~Zhu,
Prog. Part. Nucl. Phys. \textbf{107}, 237-320 (2019)
doi:10.1016/j.ppnp.2019.04.003
[arXiv:1903.11976 [hep-ph]].

\bibitem{Gell-Mann:1964ewy}
M.~Gell-Mann,
Phys. Lett. \textbf{8}, 214-215 (1964)
doi:10.1016/S0031-9163(64)92001-3

\bibitem{LHCb:2015yax}
R.~Aaij \textit{et al.} [LHCb],
Phys. Rev. Lett. \textbf{115}, 072001 (2015)
doi:10.1103/PhysRevLett.115.072001
[arXiv:1507.03414 [hep-ex]].

\bibitem{LHCb:2019kea}
R.~Aaij \textit{et al.} [LHCb],
Phys. Rev. Lett. \textbf{122}, no.22, 222001 (2019)
doi:10.1103/PhysRevLett.122.222001
[arXiv:1904.03947 [hep-ex]].

\bibitem{LHCb:2020jpq}
R.~Aaij \textit{et al.} [LHCb],
Sci. Bull. \textbf{66}, 1278-1287 (2021)
doi:10.1016/j.scib.2021.02.030
[arXiv:2012.10380 [hep-ex]].

\bibitem{LHCb:2021chn}
R.~Aaij \textit{et al.} [LHCb],
Phys. Rev. Lett. \textbf{128}, no.6, 062001 (2022)
doi:10.1103/PhysRevLett.128.062001
[arXiv:2108.04720 [hep-ex]].

\bibitem{LHCb:2022ogu}
R.~Aaij \textit{et al.} [LHCb],
Phys. Rev. Lett. \textbf{131}, no.3, 031901 (2023)
doi:10.1103/PhysRevLett.131.031901
[arXiv:2210.10346 [hep-ex]].

\bibitem{Belle:2025pey}
I.~Adachi \textit{et al.} [Belle and Belle-II],
Phys. Rev. Lett. \textbf{135}, no.4, 041901 (2025)
doi:10.1103/pf8m-6j69
[arXiv:2502.09951 [hep-ex]].


\bibitem{Karliner:2003dt}
M.~Karliner and H.~J.~Lipkin,
Phys. Lett. B \textbf{575}, 249-255 (2003)
doi:10.1016/j.physletb.2003.09.062
[arXiv:hep-ph/0402260 [hep-ph]].

\bibitem{Lebed:2015tna}
R.~F.~Lebed,
Phys. Lett. B \textbf{749}, 454-457 (2015)
doi:10.1016/j.physletb.2015.08.032
[arXiv:1507.05867 [hep-ph]].


\bibitem{Zhu:2015bba}
R.~Zhu and C.~F.~Qiao,
Phys. Lett. B \textbf{756}, 259-264 (2016)
doi:10.1016/j.physletb.2016.03.022
[arXiv:1510.08693 [hep-ph]].

\bibitem{Scoccola:2015nia}
N.~N.~Scoccola, D.~O.~Riska and M.~Rho,
Phys. Rev. D \textbf{92}, no.5, 051501 (2015)
doi:10.1103/PhysRevD.92.051501
[arXiv:1508.01172 [hep-ph]].

\bibitem{Jaffe:2003sg}
R.~L.~Jaffe and F.~Wilczek,
Phys. Rev. Lett. \textbf{91}, 232003 (2003)
doi:10.1103/PhysRevLett.91.232003
[arXiv:hep-ph/0307341 [hep-ph]].


\bibitem{Maiani:2015vwa}
L.~Maiani, A.~D.~Polosa and V.~Riquer,
Phys. Lett. B \textbf{749}, 289-291 (2015)
doi:10.1016/j.physletb.2015.08.008
[arXiv:1507.04980 [hep-ph]].

\bibitem{Wang:2026thx}
Z.~G.~Wang,
[arXiv:2606.28095 [hep-ph]].

\bibitem{Anisovich:2015cia}
V.~V.~Anisovich, M.~A.~Matveev, J.~Nyiri, A.~V.~Sarantsev and A.~N.~Semenova,
[arXiv:1507.07652 [hep-ph]].

\bibitem{Karliner:2015ina}
M.~Karliner and J.~L.~Rosner,
Phys. Rev. Lett. \textbf{115}, no.12, 122001 (2015)
doi:10.1103/PhysRevLett.115.122001
[arXiv:1506.06386 [hep-ph]].

\bibitem{Chen:2015loa}
R.~Chen, X.~Liu, X.~Q.~Li and S.~L.~Zhu,
Phys. Rev. Lett. \textbf{115}, no.13, 132002 (2015)
doi:10.1103/PhysRevLett.115.132002
[arXiv:1507.03704 [hep-ph]].

\bibitem{Ortega:2016syt}
P.~G.~Ortega, D.~R.~Entem and F.~Fern{\'a}ndez,
Phys. Lett. B \textbf{764}, 207-211 (2017)
doi:10.1016/j.physletb.2016.11.008
[arXiv:1606.06148 [hep-ph]].

\bibitem{Roca:2015dva}
L.~Roca, J.~Nieves and E.~Oset,
Phys. Rev. D \textbf{92}, no.9, 094003 (2015)
doi:10.1103/PhysRevD.92.094003
[arXiv:1507.04249 [hep-ph]].


\bibitem{Burns:2015dwa}
T.~J.~Burns,
Eur. Phys. J. A \textbf{51}, no.11, 152 (2015)
doi:10.1140/epja/i2015-15152-6
[arXiv:1509.02460 [hep-ph]].

\bibitem{Jing:2025iqs}
Z.~L.~Jing and J.~R.~Zhang,
Phys. Rev. D \textbf{112}, no.7, 074003 (2025)
doi:10.1103/xf26-q1zj
[arXiv:2507.16522 [hep-ph]].

\bibitem{Song:2025yut}
Q.~F.~Song, Q.~F.~L{\"u} and X.~Xiong,
Eur. Phys. J. C \textbf{85}, no.9, 1026 (2025)
doi:10.1140/epjc/s10052-025-14760-3
[arXiv:2501.01077 [hep-ph]].

\bibitem{Mutuk:2026zxp}
H.~Mutuk and X.~W.~Kang,
Phys. Lett. B \textbf{879}, 140650 (2026)
doi:10.1016/j.physletb.2026.140650
[arXiv:2603.27657 [hep-ph]].

\bibitem{Mironov:2015ica}
A.~Mironov and A.~Morozov,
JETP Lett. \textbf{102}, no.5, 271-273 (2015)
doi:10.7868/S0370274X15170038
[arXiv:1507.04694 [hep-ph]].


\bibitem{Takeuchi:2016ejt}
S.~Takeuchi and M.~Takizawa,
Phys. Lett. B \textbf{764}, 254-259 (2017)
doi:10.1016/j.physletb.2016.11.034
[arXiv:1608.05475 [hep-ph]].

\bibitem{Deng:2016rus}
C.~Deng, J.~Ping, H.~Huang and F.~Wang,
Phys. Rev. D \textbf{95}, no.1, 014031 (2017)
doi:10.1103/PhysRevD.95.014031
[arXiv:1608.03940 [hep-ph]].

\bibitem{Santopinto:2016pkp}
E.~Santopinto and A.~Giachino,
Phys. Rev. D \textbf{96}, no.1, 014014 (2017)
doi:10.1103/PhysRevD.96.014014
[arXiv:1604.03769 [hep-ph]].

\bibitem{Wu:2017weo}
J.~Wu, Y.~R.~Liu, K.~Chen, X.~Liu and S.~L.~Zhu,
Phys. Rev. D \textbf{95}, no.3, 034002 (2017)
doi:10.1103/PhysRevD.95.034002
[arXiv:1701.03873 [hep-ph]].

\bibitem{An:2022fvs}
H.~T.~An, S.~Q.~Luo, Z.~W.~Liu and X.~Liu,
Phys. Rev. D \textbf{105}, no.7, 074032 (2022)
doi:10.1103/PhysRevD.105.074032
[arXiv:2203.03448 [hep-ph]].

\bibitem{An:2021vwi}
H.~T.~An, K.~Chen, Z.~W.~Liu and X.~Liu,
Phys. Rev. D \textbf{103}, no.11, 114027 (2021)
doi:10.1103/PhysRevD.103.114027
[arXiv:2106.02837 [hep-ph]].

\bibitem{Genovese:1997tm}
M.~Genovese, J.~M.~Richard, F.~Stancu and S.~Pepin,
Phys. Lett. B \textbf{425}, 171-176 (1998)
doi:10.1016/S0370-2693(98)00187-7
[arXiv:hep-ph/9712452 [hep-ph]].

\bibitem{Stancu:2003if}
F.~Stancu and D.~O.~Riska,
Phys. Lett. B \textbf{575}, 242-248 (2003)
doi:10.1016/j.physletb.2003.09.061
[arXiv:hep-ph/0307010 [hep-ph]].

\bibitem{Garcilazo:2022kra}
H.~Garcilazo and A.~Valcarce,
Phys. Rev. D \textbf{105}, no.11, 114016 (2022)
doi:10.1103/PhysRevD.105.114016
[arXiv:2207.02757 [hep-ph]].


\bibitem{Regge:1959mz}
  T.~Regge,
  Nuovo Cim.\  {\bf 14}, 951 (1959)


\bibitem{Chew:1962eu}
  G.~F.~Chew and S.~C.~Frautschi,
  Phys.\ Rev.\ Lett.\  {\bf 8}, 41 (1962)


\bibitem{Collins:1971ff}
  P.~D.~B.~Collins,
  Phys.\ Rept.\  {\bf 1}, 103 (1971)


\bibitem{Irving:1977ea}
  A.~C.~Irving and R.~P.~Worden,
  Phys.\ Rept.\  {\bf 34}, 117 (1977)


\bibitem{Nambu:1978bd}
  Y.~Nambu,
  Phys.\ Lett.\ B {\bf 80}, 372 (1979)

\bibitem{Gross:2022hyw}
F.~Gross, E.~Klempt, S.~J.~Brodsky, A.~J.~Buras, V.~D.~Burkert, G.~Heinrich, K.~Jakobs, C.~A.~Meyer, K.~Orginos and M.~Strickland, \textit{et al.}
Eur. Phys. J. C \textbf{83}, 1125 (2023)
doi:10.1140/epjc/s10052-023-11949-2
[arXiv:2212.11107 [hep-ph]].


\bibitem{Inopin:1999nf}
  A.~Inopin and G.~S.~Sharov,
  Phys.\ Rev.\ D {\bf 63}, 054023 (2001).
  arXiv: hep-ph/9905499.


\bibitem{Brisudova:1999ut}
  M.~M.~Brisudova, L.~Burakovsky and T.~Goldman,
  Phys.\ Rev.\ D {\bf 61}, 054013 (2000).
  arXiv:hep-ph/9906293



\bibitem{brau:04bs}
 F.~Brau,
  Phys.\ Rev.\ D {\bf 62}, 014005 (2000).
  arXiv:hep-ph/0412170


\bibitem{Brodsky:2006uq}
  S.~J.~Brodsky,
  Eur.\ Phys.\ J.\ A {\bf 31}, 638 (2007).
  arXiv:hep-ph/0610115

\bibitem{Nielsen:2018uyn}
M.~Nielsen and S.~J.~Brodsky,
Phys. Rev. D \textbf{97}, no.11, 114001 (2018)
doi:10.1103/PhysRevD.97.114001
[arXiv:1802.09652 [hep-ph]].


\bibitem{Guo:2008he}
  X.~H.~Guo, K.~W.~Wei and X.~H.~Wu,
  Phys.\ Rev.\ D {\bf 78}, 056005 (2008).
  arXiv:hep-ph/0809.1702



\bibitem{Afonin:2014nya}
S.~S.~Afonin and I.~V.~Pusenkov,
Phys. Rev. D \textbf{90}, no.9, 094020 (2014)
[arXiv:1411.2390 [hep-ph]].


\bibitem{Sonnenschein:2018fph}
J.~Sonnenschein and D.~Weissman,
Eur. Phys. J. C \textbf{79}, no.4, 326 (2019)
[arXiv:1812.01619 [hep-ph]].


\bibitem{MartinContreras:2020cyg}
M.~A.~Martin Contreras and A.~Vega,
Phys. Rev. D \textbf{102}, no.4, 046007 (2020)
[arXiv:2004.10286 [hep-ph]].



\bibitem{MartinContreras:2023oqs}
M.~A.~Martin Contreras and A.~Vega,
Phys. Rev. D \textbf{108}, no.12, 126024 (2023)
doi:10.1103/PhysRevD.108.126024
[arXiv:2309.02905 [hep-ph]].


\bibitem{Roper:2024ovj}
L.~D.~Roper and I.~Strakovsky,
[arXiv:2410.11196 [hep-ph]].

\bibitem{Sergeenko:1994ck}
M.~N.~Sergeenko,
Z. Phys. C \textbf{64}, 315-322 (1994)
doi:10.1007/BF01557404

\bibitem{Veseli:1996gy}
S.~Veseli and M.~G.~Olsson,
Phys. Lett. B \textbf{383}, 109-115 (1996)
doi:10.1016/0370-2693(96)00721-6
[arXiv:hep-ph/9606257 [hep-ph]].

\bibitem{Selem:2006nd}
A.~Selem and F.~Wilczek,
doi:10.1142/9789812773524{\_}0030
[arXiv:hep-ph/0602128 [hep-ph]].

\bibitem{Wilczek:2004im}
F.~Wilczek,
doi:10.1142/9789812775344\_0007
[arXiv:hep-ph/0409168 [hep-ph]].



\bibitem{Burns:2010qq}
T.~J.~Burns, F.~Piccinini, A.~D.~Polosa and C.~Sabelli,
Phys. Rev. D \textbf{82}, 074003 (2010)
doi:10.1103/PhysRevD.82.074003
[arXiv:1008.0018 [hep-ph]].



\bibitem{Jakhad:2025zos}
P.~Jakhad and A.~K.~Rai,
Phys. Rev. D \textbf{113}, no.5, 054046 (2026)
doi:10.1103/bkwy-5hr9
[arXiv:2509.24772 [hep-ph]].


\bibitem{Chen:2018hnx}
J.~K.~Chen,
Eur. Phys. J. C \textbf{78}, no.3, 235 (2018)
doi:10.1140/epjc/s10052-018-5718-z


\bibitem{Chen:2018bbr}
J.~K.~Chen,
Phys. Lett. B \textbf{786}, 477-484 (2018)
doi:10.1016/j.physletb.2018.10.022
[arXiv:1807.11003 [hep-ph]].


\bibitem{Chen:2018nnr}
J.~K.~Chen,
Eur. Phys. J. C \textbf{78}, no.8, 648 (2018)
doi:10.1140/epjc/s10052-018-6134-0


\bibitem{Sindhu:2023oqo}
D.~G.~Sindhu, A.~Ranjan and H.~Nandan,
Int. J. Mod. Phys. A \textbf{38}, no.06n07, 2350044 (2023)
doi:10.1142/S0217751X23500446
[arXiv:2307.13284 [hep-ph]].


\bibitem{Ghosh:2017cck}
R.~Ghosh and A.~Bhattacharya,
Int. J. Theor. Phys. \textbf{56}, no.7, 2335-2344 (2017)
doi:10.1007/s10773-017-3386-7


\bibitem{Song:2025cla}
H.~Song, X.~R.~Liu, J.~Q.~Xie and J.~K.~Chen,
JHEP \textbf{10}, 047 (2025)
doi:10.1007/JHEP10(2025)047
[arXiv:2506.01005 [hep-ph]].


\bibitem{Feng:2023txx}
X.~Feng, J.~K.~Chen and J.~Q.~Xie,
Phys. Rev. D \textbf{108}, no.3, 034022 (2023)
doi:10.1103/PhysRevD.108.034022
[arXiv:2305.15705 [hep-ph]].


\bibitem{Liu:2026npi}
X.~R.~Liu, Q.~Liu and J.~K.~Chen,
Phys. Rev. D \textbf{114}, no.9, 094015 (2026)
doi:10.1103/24ms-nvqy
[arXiv:2606.29380 [hep-ph]].


\bibitem{Maiani:2019lpu}
L.~Maiani, A.~D.~Polosa and V.~Riquer,
Phys. Rev. D \textbf{100}, no.7, 074002 (2019)
doi:10.1103/PhysRevD.100.074002
[arXiv:1908.03244 [hep-ph]].


\bibitem{Berwein:2024ztx}
M.~Berwein, N.~Brambilla, A.~Mohapatra and A.~Vairo,
Phys. Rev. D \textbf{110}, no.9, 094040 (2024)
doi:10.1103/PhysRevD.110.094040
[arXiv:2408.04719 [hep-ph]].

\bibitem{Brodsky:2014xia}
S.~J.~Brodsky, D.~S.~Hwang and R.~F.~Lebed,
Phys. Rev. Lett. \textbf{113}, no.11, 112001 (2014)
doi:10.1103/PhysRevLett.113.112001
[arXiv:1406.7281 [hep-ph]].


\bibitem{Galkin:2023wox}
V.~O.~Galkin and E.~M.~Savchenko,
Eur. Phys. J. A \textbf{60}, no.5, 96 (2024)
doi:10.1140/epja/s10050-024-01311-9
[arXiv:2310.20247 [hep-ph]].


\bibitem{Bedolla:2019zwg}
M.~A.~Bedolla, J.~Ferretti, C.~D.~Roberts and E.~Santopinto,
Eur. Phys. J. C \textbf{80}, no.11, 1004 (2020)
doi:10.1140/epjc/s10052-020-08579-3
[arXiv:1911.00960 [hep-ph]].

\bibitem{Ferretti:2019zyh}
J.~Ferretti,
Few Body Syst. \textbf{60}, no.1, 17 (2019)
doi:10.1007/s00601-019-1483-2

\bibitem{Godfrey:1985xj}
S.~Godfrey and N.~Isgur,
Phys. Rev. D \textbf{32}, 189-231 (1985)
doi:10.1103/PhysRevD.32.189

\bibitem{Capstick:1986ter}
S.~Capstick and N.~Isgur,
Phys. Rev. D \textbf{34}, no.9, 2809-2835 (1986)
doi:10.1103/physrevd.34.2809

\bibitem{Durand:1981my}
B.~Durand and L.~Durand,
Phys. Rev. D \textbf{25}, 2312 (1982)
doi:10.1103/PhysRevD.25.2312

\bibitem{Durand:1983bg}
B.~Durand and L.~Durand,
Phys. Rev. D \textbf{30}, 1904 (1984)
doi:10.1103/PhysRevD.30.1904

\bibitem{Lichtenberg:1982jp}
D.~B.~Lichtenberg, W.~Namgung, E.~Predazzi and J.~G.~Wills,
Phys. Rev. Lett. \textbf{48}, 1653 (1982)
doi:10.1103/PhysRevLett.48.1653


\bibitem{Jacobs:1986gv}
S.~Jacobs, M.~G.~Olsson and C.~Suchyta, III,
Phys. Rev. D \textbf{33}, 3338 (1986)
[erratum: Phys. Rev. D \textbf{34}, 3536 (1986)]
doi:10.1103/PhysRevD.33.3338

\bibitem{Ferretti:2011zz}
J.~Ferretti, A.~Vassallo and E.~Santopinto,
Phys. Rev. C \textbf{83}, 065204 (2011)
doi:10.1103/PhysRevC.83.065204

\bibitem{Eichten:1974af}
E.~Eichten, K.~Gottfried, T.~Kinoshita, J.~B.~Kogut, K.~D.~Lane and T.~M.~Yan,
Phys. Rev. Lett. \textbf{34}, 369-372 (1975)
[erratum: Phys. Rev. Lett. \textbf{36}, 1276 (1976)]
doi:10.1103/PhysRevLett.34.369

\bibitem{Lucha:1991vn}
W.~Lucha, F.~F.~Schoberl and D.~Gromes,
Phys. Rept. \textbf{200}, 127-240 (1991)
doi:10.1016/0370-1573(91)90001-3


\bibitem{Gromes:1981cb}
D.~Gromes,
Z. Phys. C \textbf{11}, 147 (1981)
doi:10.1007/BF01573997

\bibitem{Chen:2022flh}
J.~K.~Chen,
Nucl. Phys. B \textbf{983}, 115911 (2022)
doi:10.1016/j.nuclphysb.2022.115911
[arXiv:2203.02981 [hep-ph]].

\bibitem{Chen:2021kfw}
J.~K.~Chen,
Eur. Phys. J. A \textbf{57}, 238 (2021)
doi:10.1140/epja/s10050-021-00502-y
[arXiv:2102.07993 [hep-ph]].


\bibitem{Chen:2023cws}
J.~K.~Chen, X.~Feng and J.~Q.~Xie,
JHEP \textbf{10}, 052 (2023)
doi:10.1007/JHEP10(2023)052
[arXiv:2308.02289 [hep-ph]].

\bibitem{Xie:2024lfo}
J.~Q.~Xie, H.~Song and J.~K.~Chen,
Eur. Phys. J. C \textbf{84}, no.10, 1048 (2024)
doi:10.1140/epjc/s10052-024-13438-6
[arXiv:2407.18280 [hep-ph]].

\bibitem{Chen:2023web}
J.~K.~Chen,
Eur. Phys. J. C \textbf{84}, no.4, 356 (2024)
doi:10.1140/epjc/s10052-024-12706-9
[arXiv:2302.06794 [hep-ph]].


\bibitem{Chen:2023djq}
J.~K.~Chen,
Nucl. Phys. A \textbf{1050}, 122927 (2024)
doi:10.1016/j.nuclphysa.2024.122927
[arXiv:2302.05926 [hep-ph]].


\bibitem{Chen:2025fyh}
J.~K.~Chen, H.~Song and X.~R.~Liu,
Eur. Phys. J. C \textbf{85}, no.11, 1344 (2025)
doi:10.1140/epjc/s10052-025-15075-z
[arXiv:2508.18899 [hep-ph]].


\bibitem{Song:2024bkj}
H.~Song, J.~Q.~Xie and J.~K.~Chen,
Eur. Phys. J. C \textbf{85}, no.5, 482 (2025)
doi:10.1140/epjc/s10052-025-14217-7
[arXiv:2408.03720 [hep-ph]].

\bibitem{Chen:2023ngj}
J.~K.~Chen, J.~Q.~Xie, X.~Feng and H.~Song,
Eur. Phys. J. C \textbf{83}, no.12, 1133 (2023)
doi:10.1140/epjc/s10052-023-12329-6
[arXiv:2310.05131 [hep-ph]].


\bibitem{Faustov:2021hjs}
R.~N.~Faustov, V.~O.~Galkin and E.~M.~Savchenko,
Universe \textbf{7}, no.4, 94 (2021)
doi:10.3390/universe7040094
[arXiv:2103.01763 [hep-ph]].


\bibitem{Xie:2024dfe}
J.~Q.~Xie, H.~Song, X.~Feng and J.~K.~Chen,
Phys. Rev. D \textbf{110}, no.7, 074039 (2024)
doi:10.1103/PhysRevD.110.074039
[arXiv:2407.04222 [hep-ph]].


\bibitem{Liu:2026vpi}
X.~R.~Liu, Q.~Liu, H.~Song and J.~K.~Chen,
Nucl. Phys. B \textbf{1031}, 117656 (2026)
doi:10.1016/j.nuclphysb.2026.117656
[arXiv:2602.16988 [hep-ph]].


\end{thebibliography}
\end{document}